\documentclass[pdftex, numberedappendix, iop, apj]{emulateapj}

\usepackage[T1]{fontenc}
\usepackage{scrextend}
\usepackage{hyperref}
\usepackage{graphicx}
\usepackage{color}
\usepackage{natbib}
\usepackage{sidecap}
\usepackage{enumitem}
\usepackage{amsmath}
\usepackage{amssymb}
\usepackage{courier}
\usepackage{xfrac}
\usepackage{sidecap}
\sidecaptionvpos{figure}{m}

\newcommand{\be}{\begin{equation}}
\newcommand{\ee}{\end{equation}}

\newcommand{\ba}{\begin{eqnarray}}
\newcommand{\ea}{\end{eqnarray}}

\newcommand{\wmap}{\textsl{WMAP}}

\newcommand{\planck}{\textsl{Planck}}

\newcommand{\lcdm}{\ensuremath{\Lambda\mathrm{CDM}}}

\shorttitle{Impact of Spectral Line Misidentification}
\shortauthors{G.~E.~Addison et al.} 

\begin{document}

\title{The Impact of Line Misidentification on Cosmological Constraints from Euclid and other Spectroscopic Galaxy Surveys}

\author{G.~E.~Addison\altaffilmark{1}, C.~L.~Bennett\altaffilmark{1}, D.~Jeong\altaffilmark{2}, E.~Komatsu\altaffilmark{3,4}, and J.~L.~Weiland\altaffilmark{1}}

\email{gaddison@jhu.edu}

\altaffiltext{1}{
Dept. of Physics \& Astronomy, The Johns Hopkins University, 3400 N. Charles St., Baltimore, MD 21218-2686
}

\altaffiltext{2}{
Department of Astronomy and Astrophysics and Institute for Gravitation and the Cosmos,
The Pennsylvania State University, University Park, PA 16802
}

\altaffiltext{3}{
Max-Planck-Institut f{\"u}r Astrophysik, Karl-Schwarzschild-Str. 1, 85741 Garching, Germany
}

\altaffiltext{4}{
Kavli Institute for the Physics and Mathematics of the Universe, Todai Institutes for Advanced Study, the University of Tokyo, Kashiwa, Japan 277-8583 (Kavli IPMU, WPI)
}

\begin{abstract}

We perform forecasts for how baryon acoustic oscillation (BAO) scale and redshift-space distortion (RSD) measurements from future spectroscopic emission line galaxy (ELG) surveys such as Euclid are degraded in the presence of spectral line misidentification. Using analytic calculations verified with mock galaxy catalogs from log-normal simulations we find that constraints are degraded in two ways, even when the interloper power spectrum is modeled correctly in the likelihood. Firstly, there is a loss of signal-to-noise ratio for the power spectrum of the target galaxies, which propagates to all cosmological constraints and increases with contamination fraction, $f_c$. Secondly, degeneracies can open up between $f_c$ and cosmological parameters. In our calculations this typically increases BAO scale uncertainties at the 10-20\% level when marginalizing over parameters determining the broadband power spectrum shape. External constraints on $f_c$, or parameters determining the shape of the power spectrum, for example from cosmic microwave background (CMB) measurements, can remove this effect. There is a near-perfect degeneracy between $f_c$ and the power spectrum amplitude for low $f_c$ values, where $f_c$ is not well determined from the contaminated sample alone. This has the potential to strongly degrade RSD constraints. The degeneracy can be broken with an external constraint on $f_c$, for example from cross-correlation with a separate galaxy sample containing the misidentified line, or deeper sub-surveys.\\

\end{abstract}

\keywords{cosmology: observations -- distance scale -- large-scale structure of universe}

\section{Introduction}

Measurements of CMB anisotropy, particularly from the \wmap\ and \planck\ satellite missions, have precisely constrained the parameters of the standard Lambda-cold dark matter (\lcdm) model and limited or ruled out many possible modifications or extensions \citep{hinshaw/etal:2013,planck/6:2018}. The next decade will see many experimental collaborations aiming to take advantage of the vast amount of cosmological information encoded in large-scale structure (LSS), building on recent galaxy clustering and weak gravitational lensing measurements \citep[e.g.,][]{alam/etal:2017,des:2018,vanuitert/etal:2018,hikage/etal:2019}. Examples include the Dark Energy Spectroscopic Instrument \citep[DESI\footnote{\href{https://www.desi.lbl.gov/}{https://www.desi.lbl.gov/}};][]{levi/etal:2013}, Euclid\footnote{\href{https://www.euclid-ec.org/}{https://www.euclid-ec.org/}} \citep{laureijs/etal:2011}, the Large Synoptic Survey Telescope \citep[LSST\footnote{\href{https://www.lsst.org/}{https://www.lsst.org/}};][]{lsstde:2012}, and Wide Field Infrared Survey Telescope \citep[WFIRST\footnote{\href{https://wfirst.gsfc.nasa.gov/}{https://wfirst.gsfc.nasa.gov/}};][]{wfirst/sdt:2015}. A key motivation for these experiments is to detect or tightly constrain deviations from cosmological constant dark energy behavior \citep[see][for review of observational methods]{weinberg/etal:2013}. Additional goals include testing General Relativity (GR) on cosmological scales (see \citeauthor{clifton/etal:2012} 2012 for a review of modified gravity theories, and, e.g., \citeauthor{jeong/schmidt:2015} 2015 for testing GR with LSS), measuring neutrino mass through the suppression of small-scale clustering \citep[e.g.,][]{font-ribera/etal:2014,boyle/komatsu:2018}, and improving on \planck's constraints on primordial non-Gaussianity \citep{planck/17:2015}.

The baryon acoustic oscillation (BAO) scale is understood to be the most robust observable in LSS clustering, and BAO measurements over a range of redshift provide valuable dark energy constraints (see Section~4 of \citeauthor{weinberg/etal:2013} 2013 for a review, and \citeauthor{alam/etal:2017} 2017, \citeauthor{bautista/etal:2017} 2017, \citeauthor{bourboux/etal:2017} 2017, \citeauthor{desbao:prep} 2017, and \citeauthor{ata/etal:2018} 2018 for the latest results). BAO measurements also tightly constrain portions of the \lcdm\ parameter space, particularly in conjunction with CMB data \citep[e.g.,][]{hinshaw/etal:2013,aubourg/etal:2015,addison/etal:2018,planck/6:2018}. They play an important role in the current Hubble constant ($H_0$) tension, providing evidence for $H_0<70$~km~s$^{-1}$~Mpc$^{-1}$ in joint fits with CMB or primordial deuterium abundance measurements within \lcdm\ \citep[e.g.,][]{addison/etal:2013b,aubourg/etal:2015,planck/13:2015,bernal/etal:2016,addison/etal:2018,desh0:2018}, while the latest local distance ladder measurement is $H_0=(73.52\pm1.62)$~km~s$^{-1}$~Mpc$^{-1}$ \citep{riess/etal:2018}.

The Baryon Oscillation Spectroscopic Survey \citep[BOSS\footnote{\href{http://www.sdss3.org/surveys/boss.php}{http://www.sdss3.org/surveys/boss.php}};][]{dawson/etal:2013} has provided ~1-2\% BAO scale measurements using luminous red galaxies (LRGs) over redshift $0.2<z<0.75$, as well as lower-precision measurements from the Lyman-$\alpha$ forest along sight-lines to quasars at $z\geq2$ \citep[see][for final Data Release 12 constraints]{alam/etal:2017,bautista/etal:2017,bourboux/etal:2017}. Euclid and WFIRST aim to `fill in' the redshift range $1<z<2$ using BAO measured from H$\alpha$ emission line galaxies (ELGs) observed using slitless spectroscopy in the near infrared. High-redshift ELGs are also targets of the extended BOSS survey \citep[eBOSS\footnote{\href{https://www.sdss.org/surveys/eboss/}{https://www.sdss.org/surveys/eboss/}};][]{dawson/etal:2016}, the Hobby-Eberly Telescope Dark Energy Experiment \citep[HETDEX\footnote{\href{http://www.hetdex.org}{http://www.hetdex.org}};][]{hill/etal:2008}, and the Subaru Prime Focus Spectrograph (PFS\footnote{\href{https://pfs.ipmu.jp/}{https://pfs.ipmu.jp/}}) cosmology survey \citep{takada/etal:2014}.

In order to sample the large cosmological volume and density of LSS tracers required for precise BAO measurements, many individual Euclid, WFIRST, and HETDEX galaxy spectra will contain only a single spectral line detected at high significance. In such cases there is a risk of line misidentification, where the detected line is not the line of interest but some other line, meaning that galaxy lies not within the target redshift range but at some (possibly very) different redshift. This causes a catastrophic redshift error with a redshift bias typically several orders of magnitude larger than the $\Delta z\sim0.001-0.01$ statistical uncertainty targeted by these experiments. The presence of such misidentified ELGs in a galaxy catalog, even at the percent level, can lead to significant bias on cosmological constraints if not accounted for \citep[hereafter P16]{pullen/etal:2016}. This is because the interloper misidentified ELGs still trace LSS and contribute clustering power from the `wrong' redshift range, making them a more serious contaminant than completely spurious catalog entries that are not ELGs at all (e.g., stars).

In this paper we perform Fisher forecasts for the extent to which cosmological constraints, particularly BAO and redshift-space distortion (RSD) measurements, are degraded for different contamination scenarios, and the extent to which this degradation can be mitigated, for example by external constraints on either the contamination fraction or portions of the cosmological parameter space. Our basic approach is to model the interloper power spectrum contribution in the multipole ELG power spectrum likelihood, with a contamination fraction that must be marginalized over. Our calculations and methodology are described in Section~2, details of the surveys we performed forecasts for are provided in Section~3, and results focusing on the Euclid [OIII] survey, where contamination is likely to be particularly severe, are presented in Section~4. In Section~5 we discuss results for other lines and surveys and identify avenues for future work. Conclusions follow in Section~6.

\section{Power spectra from catalogs containing misidentified lines}

\subsection{Simplifying assumptions}

Since we are interested in how line misidentification degrades cosmological constraints, rather than the overall constraining power or optimal analysis choices for future surveys, we make a number of simplifying assumptions:
\begin{enumerate}[label=(\roman*)]
\item We assume the ELGs are linear tracers of the linear dark matter density fluctuations. We approximately account for the impact of nonlinearity by varying a maximum cut-off scale in $k$, beyond which we assume no cosmological information can be recovered. We note that the impact of nonlinearity is relatively small at the BAO scale ($r\simeq150$~Mpc), and also smaller at the redshifts we are considering than for BAO surveys like BOSS at $z\lesssim0.7$.
\item We perform simulations and calculations for sky patches of up to $10^3$~deg$^2$, where the sky can be well-approximated as flat, and assume that constraints from larger sky areas can be obtained by simply combining the information from multiple patches. This is a reasonable approximation for scales much smaller than the patch size, including the BAO feature for ELGs at $z>0.7$, which is the focus of this work, but throws away information from scales comparable or larger than the patch size.
\item We ignore complications in the survey geometry and weighting or masking and approximate each bin in redshift as a comoving cuboid. If survey depth varies significantly with position then spatial variation in the line misidentification rate could also be introduced. An investigation of this effect for specific surveys is left to future work.
\item We neglect any time evolution within a redshift bin, for instance in matter clustering or ELG properties.
\item We assume the likelihood function for the galaxy power spectra can be approximated as Gaussian, and further that non-Gaussian contributions to the power spectrum covariance can be neglected. 
\end{enumerate}

\subsection{ELG power spectrum}

We use the redshift-space multipole power spectrum of fluctuations in the ELG overdensity as the observable that directly enters the cosmological likelihood. To forecast constraints we therefore need to compute the mean and covariance of the multipole power spectrum as a function of cosmological and ELG parameters.

In Appendix~A we connect the ELG power spectrum to the underlying density fluctuations and compile some analytic results from the literature for the multipoles of the linear theory redshift space power spectrum and the covariance between different multipoles. Given the Fourier coefficients of the galaxy density field at wavevector ${\bf k}$, $\delta_g({\bf k})$, we write down an estimator for the ELG multipole power spectrum at multipole $\ell$ as
\be
\hat{P}_{g,\ell}(k)=\frac{2\ell+1}{2}\int_{-1}^1d\mu_{\bf k}\,\delta_g({\bf k})\delta^*_g({\bf k})\mathcal{P}_{\ell}(\mu_{\bf k})-\frac{\delta_{\ell0}}{n_g},
\ee
where $\mu_{\bf k}=k_{\parallel}/|{\bf k}|$ is the cosine of the angle between ${\bf k}$ and the line of sight, $\mathcal{P}_{\ell}$ is a Legendre polynomial, and $n_g$ is the number density of galaxies, so that $1/n_g$ is the shot-noise contribution to the power spectrum, which is subtracted for the monopole, $\ell=0$ ($\delta_{\ell0}$ here is the Kronecker delta). The mean and covariance of this estimator can be computed analytically for linear theory redshift-space distortions \citep{kaiser:1987}, and are non-zero only for $\ell=0,2,4$. Full expressions are provided in Appendix~A. The mean of the monopole, for example, is given by
\be
\left\langle\hat{P}_{g,\ell=0}(k)\right\rangle=\left(1+\frac{2}{3}\beta+\frac{1}{5}\beta^2\right)b_g^2P_m(k),
\ee
where $\beta=f/b_g$, $f$ is the derivative of the cosmological growth rate, $b_g$ is the galaxy bias, and $P_m(k)$ is the linear matter power spectrum.

\subsection{Adding interlopers with misidentified lines}

We follow the process described by P16 and \cite{leung/etal:2017} for adding interloper galaxies with misidentified spectral lines. The observed emission line wavelength $\lambda$ is related to the rest-frame wavelength $\lambda_0$ by $\lambda=\lambda_0(1+z)$. A misidentified line with rest-frame wavelength $\lambda_{\rm int}$ is observed at redshift $z_{\rm int}$ such that $\lambda=\lambda_{\rm int}(1+z_{\rm int})$. When angular coordinates and redshift are transformed to three-dimensional coordinates in order to estimate the ELG power spectrum the interloper coordinates are calculated incorrectly. The coordinates are also remapped anisotropically, because transverse separations scale like the proper motion distance\footnote{Also referred to as the comoving angular diameter distance. We follow recent BAO literature in using the $D_M$ notation \citep[e.g.,][]{alam/etal:2017}.}, $D_M(z)$, while line-of-sight separations scale like $D_H(z)=(1+z)c/H(z)$. We follow P16 introducing transverse and line-of-sight remapping parameters, $\gamma_{\bot}$ and $\gamma_{\parallel}$, given by
\be
\begin{split}
\gamma_{\bot}&=\frac{D_M(z)}{D_M(z_{\rm int})}\\
\gamma_{\parallel}&=\frac{D_H(z)}{D_H(z_{\rm int})}=\frac{(1+z)/H(z)}{(1+z_{\rm int})/H(z_{\rm int})}=\frac{\lambda_{\rm int}H(z_{\rm int})}{\lambda_0 H(z)}.
\end{split}
\ee
and a contamination fraction, $f_c$, such that $f_c$ is the fraction of the total number of galaxies in the catalog where line misidentification has occurred. Writing the total number density as $n_t=n_g+n_{\rm int}$, with the subscripts `t', `g', and `int' denoting total, target galaxy, and interloper, respectively, the number density of interlopers is
\be
n_{\rm int}=f_cn_t=\frac{f_c}{1-f_c}n_g.
\ee
Note that $n_{\rm int}$ here is calculated using the target ELG survey volume. The ELG overdensity in the contaminated catalog is
\be
\delta_t({\bf x})=(1-f_c)\delta_g({\bf x})+f_c\delta_{\rm int}({\bf x}_{\bot}/\gamma_{\bot},{\bf x}_{\parallel}/\gamma_{\parallel}),
\ee
where ${\bf x}$ is the three-dimensional position vector, and the volume integral in the Fourier transform picks up a factor $\gamma_{\bot}^2\gamma_{\parallel}$, so we have
\be
\delta_t({\bf k})=(1-f_c)\delta_g({\bf k})+f_c\gamma_{\bot}^2\gamma_{\parallel}\delta_{\rm int}(\gamma_{\bot}{\bf k_{\bot}},\gamma_{\parallel}{\bf k_{\parallel}}).
\ee
One factor of $\gamma_{\bot}^2\gamma_{\parallel}$ is used remapping the coordinates of the Dirac delta in the covariance of $\delta_{\rm int}({\bf k})$ (equation 8 of P16):
\be
\left\langle\delta_{\rm int}({\bf k})\delta^*_{\rm int}({\bf k'})\right\rangle=\gamma_{\bot}^2\gamma_{\parallel}\delta_D^3({\bf k}-{\bf k'})P_{\rm int}(\gamma_{\bot}{\bf k}_{\bot},\gamma_{\parallel}{\bf k}_{\parallel}).
\ee
Assuming there is no correlation between the target and interloper populations (i.e., no redshift overlap), we can calculate the mean of the estimator in equation (1) for the contaminated case :
\be
\left\langle\hat{P}_{t,\ell}(k)\right\rangle=(1-f_c)^2P_{g,\ell}(k)+f_c^2\gamma_{\bot}^2\gamma_{\parallel}P_{\rm int, \ell}(\gamma_{\bot}{\bf k}_{\bot},\gamma_{\parallel}{\bf k}_{\parallel}),
\ee
where
\be
\begin{split}
P_{\rm int, \ell}&(\gamma_{\bot}{\bf k}_{\bot},\gamma_{\parallel}{\bf k}_{\parallel})\\&=\frac{2\ell+1}{2}\int_{-1}^1d\mu_{\bf k}P_{\rm int}(\gamma_{\bot}{\bf k}_{\bot},\gamma_{\parallel}{\bf k}_{\parallel})\mathcal{P}_{\ell}(\mu_{\bf k})
\end{split}
\ee
and
\be
\begin{split}
P_{\rm int}&(\gamma_{\bot}{\bf k}_{\bot},\gamma_{\parallel}{\bf k}_{\parallel})\\&=(1+\beta_{\rm int}\mu^2_{{\bf k}_{\rm int}})^2b_{\rm int}^2P_m\left(k=\sqrt{\gamma^2_{\bot}k^2_{\bot}+\gamma^2_{\parallel}k^2_{\parallel}},z=z_{\rm int}\right),
\end{split}
\ee
with
\be
\mu_{{\bf k}_{\rm int}}=\frac{\gamma_{\parallel}k_{\parallel}}{\sqrt{\gamma^2_{\bot}k^2_{\bot}+\gamma^2_{\parallel}k^2_{\parallel}}}.
\ee

Expressions for the covariance $\left\langle\left(\hat{P}_{\ell,t}(k)-\left\langle\hat{P}_{\ell,t}(k)\right\rangle\right)\left(\hat{P}_{\ell',t}(k')-\left\langle\hat{P}_{\ell',t}(k')\right\rangle\right)\right\rangle$ can be derived in an analogous way to equation (A8), although there are no closed-form expressions analogous to (A9). There are separate contributions from the target galaxy sample variance, interloper sample variance, and shot-noise, as well as cross-terms. Note that the strong scaling of the covariance contributions with $f_c$ (e.g., the target ELG galaxy sample variance scales as $(1-f_c)^4$) means approximating the contaminated covariance with the pure target ELG covariance may be a poor approximation unless $f_c\ll1$.

For models close to \lcdm, $D_M$ increases monotonically with redshift. The transverse remapping parameter $\gamma_{\bot}$ is thus greater than one for lower-redshift interlopers and less than one for higher-redshift interlopers. The quantity $(1+z)/H(z)$ increases at low redshift but peaks at $z\simeq0.7$ before decreasing, eventually falling off like $(1+z)^{-1/2}$ at redshifts where the universe is essentially completely matter dominated and the dark energy density is negligible. For the target and contaminant lines relevant to Euclid, HETDEX, or WFIRST, $\gamma_{\parallel}$ differs from unity only at the 10-15\% level, while $\gamma_{\bot}$ can vary by an order of magnitude when the difference between target and contaminant redshifts is large.

In the simple but unrealistic case where $\gamma_{\bot}=\gamma_{\parallel}=\gamma$ the interloper coordinate remapping is isotropic and when we measure power at wavenumber $k$ we are in fact measuring interloper power from wavenumber $\gamma k$ instead. In other words, the measured power spectrum at each multipole contains a `squashed' or `stretched' contribution from the interloper power at that same multipole.

The realistic anisotropic case where $\gamma_{\bot}\neq\gamma_{\parallel}$ is more complicated. The power at wavenumber $k$ in the contaminated sample contains contributions from a range of scales in the interloper spectrum (scales between $\gamma_{\parallel}k$ and $\gamma_{\bot}k$) and the integral in equation (9) no longer has a closed-form solution. The anisotropy causes power to be transferred between multipoles. For example, interlopers at higher redshift than the target ELGs have their quadrupole and hexadecapole power enhanced relative to the monopole. The effects of coordinate remapping are discussed in more detail for specific combinations of target and interloper lines in Sections~4 and 5.

\subsection{Approximations for finite volume}

In practice when we are considering a finite survey volume only discrete $k$-modes are available, and the expressions in Sections~2.2 and 2.3 need to be modified to account for this. We work with bins in $|{\bf k}|$ and write down an estimator for the binned power spectrum as a weighted sum over ${\bf k}$-modes:
\be
\hat{P}_{t,\ell}(b)=\frac{2\ell+1}{N_b}\sum_{{\bf k}\in b}\delta_t({\bf k})\delta^*_t({\bf k})\mathcal{P}_{\ell}(\mu_{{\bf k}})-\frac{\delta_{\ell0}}{n_t}
\ee
where $N_b$ is the number of independent modes in bin $b$ and the sum runs over these modes. Provided the power spectrum does not vary significantly over the modes within a given bin $b$ we can make the approximation that
\be
\langle\hat{P}_{t,\ell}(b)\rangle\simeq P_{t,\ell}(k_b),
\ee
where $k_b$ is the central wavenumber in the bin. The expressions for the covariance of the estimator (equations A8 and A9), and their analogs for the interlopers, are similarly evaluated at $k_b$, and are multiplied by a factor $2/N_b$, with the factor of two arising because of only counting independent modes (reality of the density field means only half the Fourier modes are independent).

A second complication is that the integrals over $\mu_k$ in equations (1) and (9) should be replaced by a sum over the discrete set of $\mu_{\bf k}$ values corresponding to the ${\bf k}$-modes falling in each bin. To simplify calculations for different survey volumes we continue to use the integrals and ignore the exact configuration of modes. This is a reasonable approximation provided there are enough modes in each bin to provide roughly uniform coverage in $\mu_{\bf k}$. 

We compared our calculations to results obtained from mock galaxy catalogs from redshift-space log-normal simulations of the cosmological density field. This is an important check of both the finite volume approximations and results when interlopers are included and integrals no longer have closed-form solutions. The method and code to generate the simulations are described by \cite{agrawal/etal:2017}. More details are provided in Appendix~B. We also use this code to compute $N_b$ for each bin and survey.

\subsection{Forecasting methodology}

Our main goal is to forecast how cosmological parameter constraints are degraded in different interloper scenarios, and examine the extent to which the impact of interlopers can be mitigated. We emphasize that this is different from the approach described in Section~2 of P16, where the focus is on estimating the \emph{bias} in cosmological parameters when the interlopers are present but not accounted for in the fitting.

The steps in our calculations are as follows: (i) choose fiducial values of cosmological parameters and parameters relating to the ELG and interloper populations (e.g., $f_c$, $b_g$), (ii) specify a redshift range for the target ELGs and calculate the dimensions of the comoving volumes (cuboids) containing the target and interloper ELGs, as well as the number of Fourier modes in each $k$ bin, (iii) calculate the mean and covariance of the estimator defined in equation (12), including target and interloper ELGs, as described above, and (iv) calculate the Fisher matrix, $\mathcal{F}$, for parameters of interest (including nuisance parameters like galaxy bias) by computing numerical derivatives of the multipole power spectra with respect to the parameters:
\be
\mathcal{F}_{ij}=\sum_{\ell,\ell'}\sum_b\frac{\partial P_{t,\ell}(k_b)}{\partial\theta_i}\left[\mathcal{C}_{\ell,\ell'}(k_b,k_b)\right]^{-1}\frac{\partial P_{t,\ell'}(k_b)}{\partial\theta_j}.
\ee
Note that we include correlations between multipoles but not $k$ bins as described in Appendix~B. The sample variance contribution to the covariance $\mathcal{C}_{\ell,\ell'}(k_b,k_b)$ also depends on the cosmological and ELG parameters. We found that this dependence impacts results for the Euclid [OIII] sample in Section~4 at the percent level and so neglect it, evaluating the covariance only for the fiducial parameters. See, for example, \cite{heavens:2009} or \cite{verde:2010} for more discussion of Fisher matrices in the context of cosmological analysis.

To facilitate calculating numerical derivatives and removing or rescaling the BAO `wiggles' in the power spectrum we calculated linear matter power spectra using code from the same package used to generate log-normal simulations in Appendix~B. The code implements the approximations and fitting functions described by \cite{eisenstein/hu:1998}. This is less accurate than the matter power spectrum produced by Code for Anisotropies in the Microwave Background \citep[\texttt{CAMB};][]{lewis/etal:2000} but adequate to assess the loss of information in the presence of interlopers. Our results were calculated assuming a cosmology with $\{\Omega_b,\Omega_m,h,n_s,\sigma_8\}=\{0.0456,0.274,0.704,0.963,0.809\}$, based on \wmap\ analysis \citep{komatsu/etal:2011}. Despite the precision of future spectroscopic surveys, changing the input parameters, for example using more recent constraints from \planck, does not significantly impact our findings, which are largely based on the comparison between contaminated and pure ELG samples rather than overall constraining power.

We performed calculations with bin width $\Delta k=0.005$~$h$Mpc$^{-1}$ and used bin centers covering $k_{\rm min}=0.005$~$h$Mpc$^{-1}$ to $k_{\rm max}=0.3$~$h$Mpc$^{-1}$. The choice of these bounds is fairly arbitrary. Arguably, the upper limit should depend on the redshift of the target ELGs since the impact of nonlinearity is redshift dependent, for example. In Section~4.5 we show that even large changes to the range of scales do not change our main conclusions regarding the effect of interlopers on BAO and RSD constraints.

\subsection{Recovery of isotropic BAO scale}

The BAO scale imprinted at recombination is a key observable in large-scale structure clustering and an important driver for the ELG number density and volume surveyed in current and future surveys. A range of methods have been developed to robustly extract the BAO scale \citep[e.g.,][]{eisenstein/etal:2005,sanchez/etal:2008,beutler/etal:2011,padmanabhan/etal:2012,sanchez/etal:2013,kazin/etal:2014}. Our approach is motivated by the method used in the multipole power spectrum analysis of the final BOSS release \citep[Sections~7.2 and 7.3 of][see references in that paper for earlier work]{beutler/etal:2017}. We consider a shift in the location of the BAO `wiggles' rather than a shift in the full power spectrum, however, since we are also interested in the effect of marginalizing over parameters determining the broadband shape.

We consider a shift in the apparent position of the BAO peak in the galaxy correlation function relative to a fiducial cosmological model of $r_{\rm BAO}\to\alpha r_{\rm BAO}$. If the fiducial cosmological model is correct then $\alpha=1$, with other values indicating a difference between the data and the fiducial model in either the conversion of ELG angular position and redshift to three-dimensional coordinates, or the absolute sound horizon at decoupling, $r_d$. Following \cite{eisenstein/etal:2005}, we have
\be
\alpha=\frac{\left[D_V(z)/r_d\right]}{\left[D_V(z)/r_d\right]_{\rm fid}},
\ee
where $D_V(z)$ is an angle-averaged combination of $D_M(z)$ and $1/H(z)$:
\be
D_V(z)=\left[D_M^2(z)\frac{cz}{H(z)}\right]^{1/3}.
\ee
To apply the shift in the BAO peaks in Fourier space we write the linear matter power spectrum as
\be
P_m(k)=P_{nw}(k)+P_w(k),
\ee
where $P_{nw}$ is the `no wiggles' power spectrum computed using a transfer function without the baryonic oscillatory features \citep[equation 30 of][]{eisenstein/hu:1998}, and $P_w$ is the power spectrum of the `wiggles' (i.e., BAO). We then have
\be
\begin{split}
P_m(k,\alpha)&=P_{nw}(k)+\frac{1}{\alpha^3}P_w(k/\alpha)\\
&=P_{nw}(k)+\frac{1}{\alpha^3}\left[P_m(k/\alpha)-P_{nw}(k/\alpha)\right].
\end{split}
\ee
We forecast constraints on the isotropic BAO scale by treating $\alpha$ as a model parameter and numerically computing derivatives using equation (14). Note that we only consider a shift in the BAO scale for the target ELGs and do not attempt to use the interloper power to constrain the BAO or other cosmological parameters.

\subsection{Recovery of anisotropic BAO scale}

Current state-of-the-art BAO surveys like BOSS have the constraining power to measure the BAO scale along the line of sight and in the transverse direction simultaneously instead of simply an angle-averaged isotropic scale \citep[e.g.,][]{anderson/etal:2014,font-ribera/etal:2014b,alam/etal:2017}. As already discussed in the context of interloper coordinate remapping, line-of-sight separations scale with $(1+z)/H(z)$ while transverse separations scale with $D_M$. Separate constraints on these quantities from anisotropic BAO contain additional cosmological information over a single angle-averaged measurement \citep[e.g., Section~3.2 of][]{addison/etal:2018}. To investigate how anisotropic BAO constraints from future surveys are impacted by interlopers we introduce separate transverse and line-of-sight BAO dilation parameters, $\alpha_{\bot}$ and $\alpha_{\parallel}$, so that
\be
\begin{split}
P_m(&k_{\bot},k_{\parallel}, \alpha_{\bot}, \alpha_{\parallel})=P_{nw}(k_{\bot},k_{\parallel})\\&+\frac{1}{\alpha_{\bot}^2\alpha_{\parallel}}[P_m(k_{\bot}/\alpha_{\bot},k_{\parallel}/\alpha_{\parallel})-P_{nw}(k_{\bot}/\alpha_{\bot},k_{\parallel}/\alpha_{\parallel})].
\end{split}
\ee

As in the isotropic case above, departure from unity in these parameters indicates that the data prefer either a different conversion from angles and redshifts to three-dimensional coordinates, or a different absolute sound horizon scale, compared to the fiducial model. Specifically, we have
\be
\begin{split}
\alpha_{\bot}&=\frac{D_M(z)/r_d}{\left[D_M(z)/r_d\right]_{\rm fid}}\\
\alpha_{\parallel}&=\frac{D_H(z)/r_d}{\left[D_H(z)/r_d\right]_{\rm fid}}=\frac{\left[H(z)r_d\right]_{\rm fid}}{H(z)r_d},\\
\end{split}
\ee
where $D_M(z)$ and $D_H(z)$ are defined in Section~2.3. We forecast constraints on the anisotropic BAO scale measurements by treating $\alpha_{\bot}$ and $\alpha_{\parallel}$ as additional model parameters and numerically computing derivatives using equation (14). Note that $\alpha_{\bot}$ and $\alpha_{\parallel}$ are always varied together as a pair.

\subsection{Measuring growth of structure with redshift-space distortions (RSD)}

There is a complete degeneracy between the linear galaxy bias, $b_g$, and the amplitude of the matter power spectrum, $P_m$, from the monopole power spectrum alone (equation 2). Adding the quadrupole provides a constraint on $\beta=f/b_g$ (equation A6). Since the matter power spectrum amplitude is proportional to $\sigma^2_8(z)$, the constraints on $\beta$ and $b_g^2\sigma^2_8(z)$ can be combined to produce a constraint on the quantity $f\sigma_8(z)$, removing the dependence on the bias. This is the approach used to obtain cosmological constraints from RSD in recent surveys \citep[e.g.,][]{beutler/etal:2012,howlett/etal:2015,pezzotta/etal:2017,alam/etal:2017}. It is well suited to measurements over a modest range of scales, where a fiducial model for the shape of the power spectrum can be assumed without significantly impacting results.

For the RSD Fisher forecasts discussed in Sections~4 and 5 we did not assume a fiducial power spectrum shape and instead varied $b_g$, $\sigma_8$, and $\Omega_m$ (in \lcdm, $\Omega_m$ determines $f$) separately, along with additional parameters like $h$ and $n_s$ that determine $P_m(k)$. We also investigated holding the power spectrum shape fixed and varying $\beta$ and $b_g\sigma_8(z)$, as described above, and did not find any qualitatively different behavior.

\begin{table*}
   \centering
   \caption{Survey properties assumed for the calculations in this work}
   \begin{tabular}{llll}
\hline
\hline
Experiment&Euclid&Euclid&HETDEX\\
\hline
Target line&[OIII] 5007+4959\AA&H$\alpha$ 6563\AA&Ly$\alpha$ 1216\AA\\
Redshift range&$1.5<z<2.3$&$0.9<z<1.5$&$1.9<z<3.5$\\
Effective redshift&1.9&1.2&2.7\\
ELG bias, $b_g$&1.7&1.5&2.0\\
Surface density, $n_g$ [deg$^{-2}$]&282&3900&2800\\
Interloper line&H$\alpha$ 6563\AA&[OIII] 5007+4959\AA&[OII] 3726+3729\AA\\
Interloper redshift range&$0.9<z<1.5$&$1.5<z<2.3$&$0<z<0.5$\\
Interloper effective redshift&1.2&1.9&0.2\\
Interloper ELG bias, $b_{\rm int}$&1.5&1.7&1.0\\
Interloper surface density, $n_{\rm int}$ [deg$^{-2}$]&3900&282&3333\\
Transverse remapping factor, $\gamma_{\bot}$&1.3&0.7&7.3\\
Line-of-sight remapping factor, $\gamma_{\parallel}$&0.9&1.1&0.9\\
Volume remapping factor, $\gamma_{\bot}^2\gamma_{\parallel}$&1.7&0.6&47.6\\
\hline
\end{tabular}

{\bf Notes.} The $\gamma$ remapping factors are defined in equation (3) of Section~2.3. The two Euclid columns correspond to the two target lines, [OIII] and H$\alpha$, respectively.
\end{table*}

\subsection{Power spectrum signal-to-noise ratio}

In addition to considering the effect of interlopers on cosmological parameter determination we also found it helpful to examine their impact at the power spectrum level. We define an overall power spectrum signal-to-noise ratio, $\mathcal{S}$, where here `signal' is the power spectrum of the target ELGs, by summing over all the multipoles and power spectrum bins,
\be
\mathcal{S}^2=\sum_{\ell\ell'}\sum_b(1-f_c)^4P_{g,\ell}(k_b)\cdot\left[\mathcal{C}_{\ell\ell'}(k_b,k_b)\right]^{-1}\cdot P_{g,\ell'}(k_b),
\ee
where $k_b$ denotes the $k$-modes in bin $b$ and we take the bins as independent (Appendix~B). The covariance $\mathcal{C}$ includes sample variance and shot-noise for both the target and interloper ELGs.

The value of $\mathcal{S}$ corresponds to the significance at which the target ELG power spectrum is measured to be non-zero, assuming perfect knowledge of the target and interloper power spectra, and $f_c$. It is equivalent to the Fisher uncertainty on the overall power spectrum amplitude (proportional to $\sigma_8^2$) while keeping all other parameters fixed. Comparing the Fisher forecasts to $\mathcal{S}$ can be useful for assessing whether cosmological constraints are degraded in the presence of interlopers due to the loss of information in the power spectrum, or some additional parameter degeneracy opening up, for example with $f_c$. We note, however, that the loss of power spectrum signal-to-noise defined in this way does not represent a strict lower bound on the degradation of BAO or other parameter uncertainties.

\section{Experiment properties}

Properties for the surveys and emission lines we consider are listed in Table~1 and discussed in more detail below. We show results in Section~4 below for the Euclid [OIII] survey contaminated by H$\alpha$, motivated by the fact that severe contamination is possible in this case. Figure~15 of P16 shows that the WFIRST [OIII] survey may have H$\alpha$ contamination at the level of tens of percent, even after using secondary line identification. The problem is likely to be more severe for Euclid given the lower signal-to-noise line detection threshold. We discuss our conclusions regarding recovery of BAO and RSD information as a function of contamination fraction in the context of other surveys and target lines in Section~5.

One of the strengths of spectroscopic redshift surveys for dark energy constraints is being able to make BAO and RSD measurements in narrow, possibly overlapping redshift bins \citep[e.g.,][]{wang/etal:2017}. In this work we focus on comparing cosmological constraints from contaminated ELG samples including misidentified lines with corresponding constraints from pure samples, in order to directly assess the impact of interlopers. As a result, the exact choices of redshift binning, redshift range, or effective ELG bias do not significantly impact our conclusions, and for simplicity we show results without subdividing ELG samples by redshift.

\subsection{Euclid}
The Euclid mission design includes a spectroscopic galaxy survey over around 15000~deg$^2$ using its Near Infrared Spectrometer and Photometer (NISP) instrument \citep{laureijs/etal:2011}. Its red grisms cover $1250-1850$~nm, corresponding to a redshift range of roughly $0.9<z<1.8$ for the primary line targeted in the survey, H$\alpha$ (6563\AA), with a blue grism covering shorter wavelengths. A secondary cosmological target is the [OIII] doublet at 5007 and 4959\AA, which is observed in the red grism for $1.5<z<2.7$. Other lines, including Ly$\alpha$, [OII], H$\beta$, and [SII], from ELGs at other redshifts will also fall into the red grism wavelength range. These lines may also be misidentified as H$\alpha$ or [OIII], depending on how much additional information (for instance from equivalent widths, or photometry) is brought to bear when constructing ELG catalogs (P16). Note that the NISP has the resolution to resolve the [OIII] doublet, but the 4959 line flux is only a third of the 5007 flux, meaning that for low signal-to-noise spectra a noise fluctuation may either render the 4959 line undetectable or create a false doublet when the detected line is actually H$\alpha$.

The H$\alpha$ source density of 3900~deg$^{-2}$ in Table~1 is taken from the lower range of recent forecasts by \cite{merson/etal:2018} for H$\alpha+$[NII] blended flux limit of $2\times10^{-16}$~erg~s$^{-1}$~cm$^{-2}$. The [OIII] number density of 282~deg$^{-2}$ is from predictions from the Hubble Space Telescope Wide Field Camera 3 Infrared Spectroscopic Parallels (WISP) program \citep{colbert/etal:2013} and a flux limit of around $3\times10^{-16}$~erg~s$^{-1}$~cm$^{-2}$. To match the predictions from \cite{colbert/etal:2013} we restrict the redshift ranges to those shown in Table~1, and do not include high-redshift H$\alpha$ ELGs at $1.5<z<1.8$ or [OIII] ELGs at $2.3<z<2.7$. There are substantial uncertainties in source density predictions and we examine the implications of large changes in these values in Section~5.1.

\subsection{HETDEX}
The HETDEX survey is designed to observe 840,000 Ly$\alpha$-emitting galaxies (LAEs) over $1.9<z<3.5$ ($3500-5500$\AA) in a 300~deg$^2$ field \citep{hill/etal:2008}. An additional smaller field will also be observed, however we perform calculations for the main field only, following \cite{leung/etal:2017}. The main interloper line is the [OII] doublet around 3727\AA, which will not be resolved with the HETDEX spectrograph. The interloper ELGs in this case are at $z<0.5$, much lower redshift than the target lines, which leads to $\gamma_{\bot}\gg1$ and a more pronounced anisotropic coordinate remapping than for the other surveys and lines we consider. \cite{leung/etal:2017} forecast a fractional contamination of the HETDEX LAE sample by [OII] of up to few percent based on a Bayesian classification scheme using equivalent width distributions. Note that there will be no contamination for LAEs at $z<2.065$ because this would require observing [OII] at wavelengths shorter than $3727$\AA\ (i.e., a blueshift). We follow \cite{leung/etal:2017} and assign a linear bias $b_g=2.0$ for the LAEs, and $b_{\rm int}=1.0$ for the low-redshift [OII] interlopers.

\subsection{WFIRST}
The planned WFIRST high-latitude spectroscopic survey covers 2227~deg$^2$ and will target the same emission lines as Euclid, H$\alpha$ at $1.06<z<1.88$ and [OIII] at $1.88<z<2.77$ \citep[Section~2.2.4 of the Science Definition Team, SDT, report,][]{wfirst/sdt:2015}. The forecast source densities in the SDT report are around 7400~deg$^{-2}$ for H$\alpha$ and 600~deg$^{-2}$ for [OIII], although \cite{merson/etal:2018} forecast a higher H$\alpha$ density of $10400-15200$~deg$^{-2}$ for the same flux cut of $1\times10^{-16}$~erg~s$^{-1}$~cm$^{-2}$ including blended H$\alpha+$[NII] flux. The impact of H$\alpha$-[OIII] line misidentification on WFIRST cosmological constraints is discussed in Section~5.1.

\section{Results}

\begin{figure*}
\hspace{-0.6cm}
\includegraphics{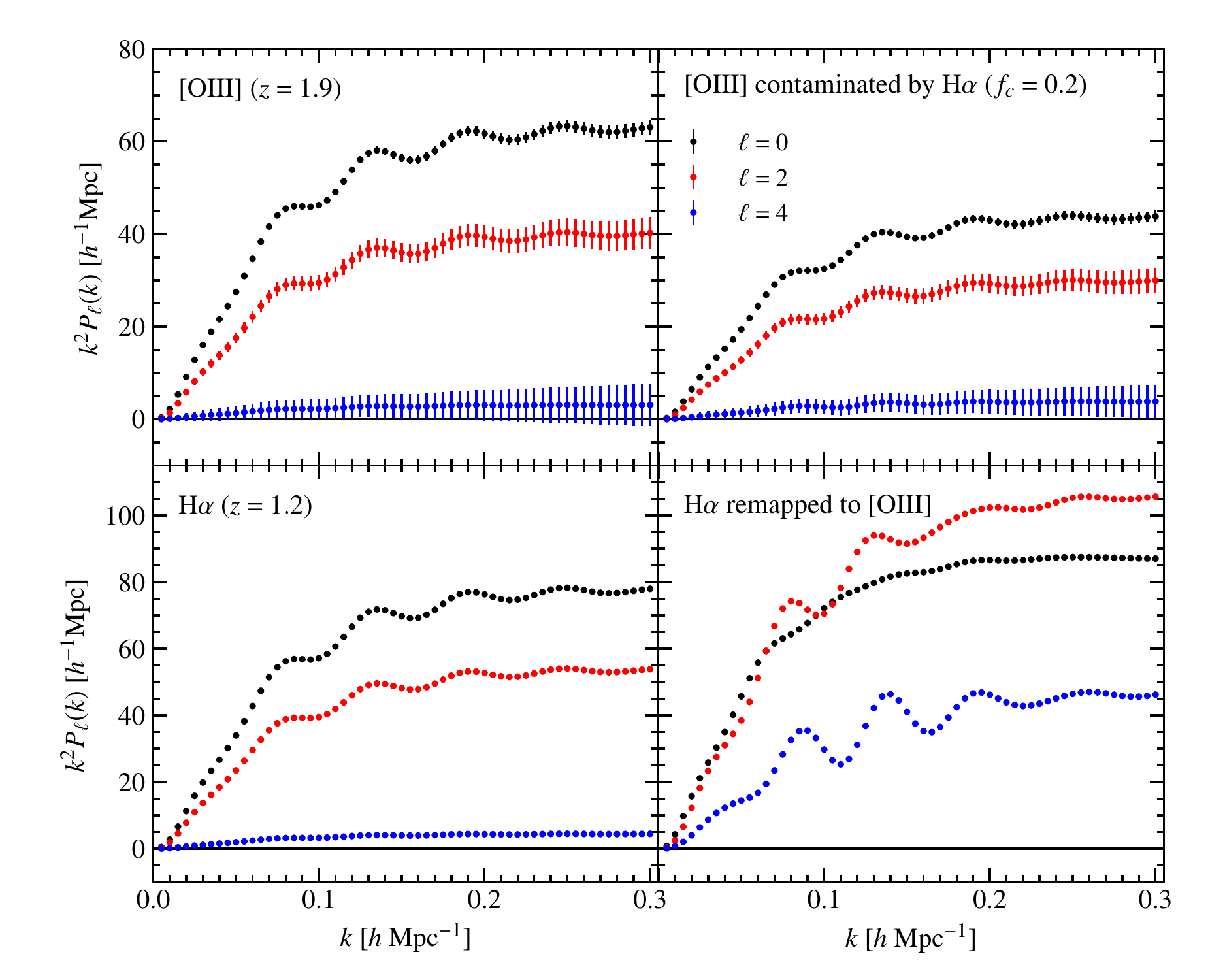}
\caption{Multipole power spectra forecasts for Euclid [OIII] ELGs, calculated for a 15000~deg$^2$ survey by combining constraints from 600~deg$^2$ patches. All quantities are in comoving coordinates. Error bars are $1\sigma$ errors for bins of width $\Delta k=0.005$~$h$Mpc$^{-1}$ and include contributions from sample variance and shot-noise. \emph{Top left:} Power spectrum of the target [OIII] ELGs in the absence of interlopers. \emph{Top right:} Power spectrum of [OIII] ELGs contaminated with interloper H$\alpha$ ELGs for a contamination fraction $f_c=0.2$. The interlopers contribute anisotropic power and suppress the monopole. \emph{Bottom left:} Shape of H$\alpha$ ELG power spectra without coordinate remapping. Differences in overall and relative amplitudes of the different multipoles compared to [OIII] are due to differences in galaxy bias (1.7 for [OIII], 1.5 for H$\alpha$) and growth of structure between $z=1.9$ and 1.2. \emph{Bottom right:} Shape of H$\alpha$ ELG power spectra for galaxies that are misidentified as [OIII] and have coordinates remapped. The quadrupole and hexadecapole are enhanced relative to the monopole for lower-redshift interlopers. Each $k$ bin in the [OIII] coordinates receives contributions from a range of $k$ in the true H$\alpha$ coordinates, causing a smearing out of BAO wiggles in the monopole power spectrum.}
\end{figure*}

\subsection{Contaminated power spectra}

The top left panel of Figure~1 shows a forecast of the monopole, quadrupole, and hexadecapole power from [OIII] ELGs at $z=1.9$ for a 15000~deg$^2$ Euclid galaxy survey (where, as stated earlier, we approximate the constraining power of the full survey by imagining combining separate constraints from 600~deg$^2$ patches). The top right panel shows the power spectrum from the same [OIII] ELGs with the addition of misidentified H$\alpha$ ELGs from $z=1.2$ for a fractional contamination of $f_c=0.2$. The bottom panels of Figure~1 show the H$\alpha$ power spectrum in the correct coordinates and when misidentified as [OIII]. The anisotropic interloper remapping causes a suppression of the interloper monopole and enhancement of the quadrupole and hexadecapole. Whether this causes a net suppression or enhancement in the contaminated power spectrum (top right panel) depends on $f_c$. If $f_c$ is small, the main effect is a suppression of power in every multipole that scales like $(1-f_c)^2$, from the first term of equation (8). In Figure~1, $f_c=0.2$ is large enough that an enhancement of the contaminated hexadecapole relative to the monopole is apparent by eye. The addition of the interloper power in the second term in equation (8) has roughly compensated the $(1-f_c)^2$ loss.

\begin{figure*}
\hspace{-0.6cm}
\includegraphics{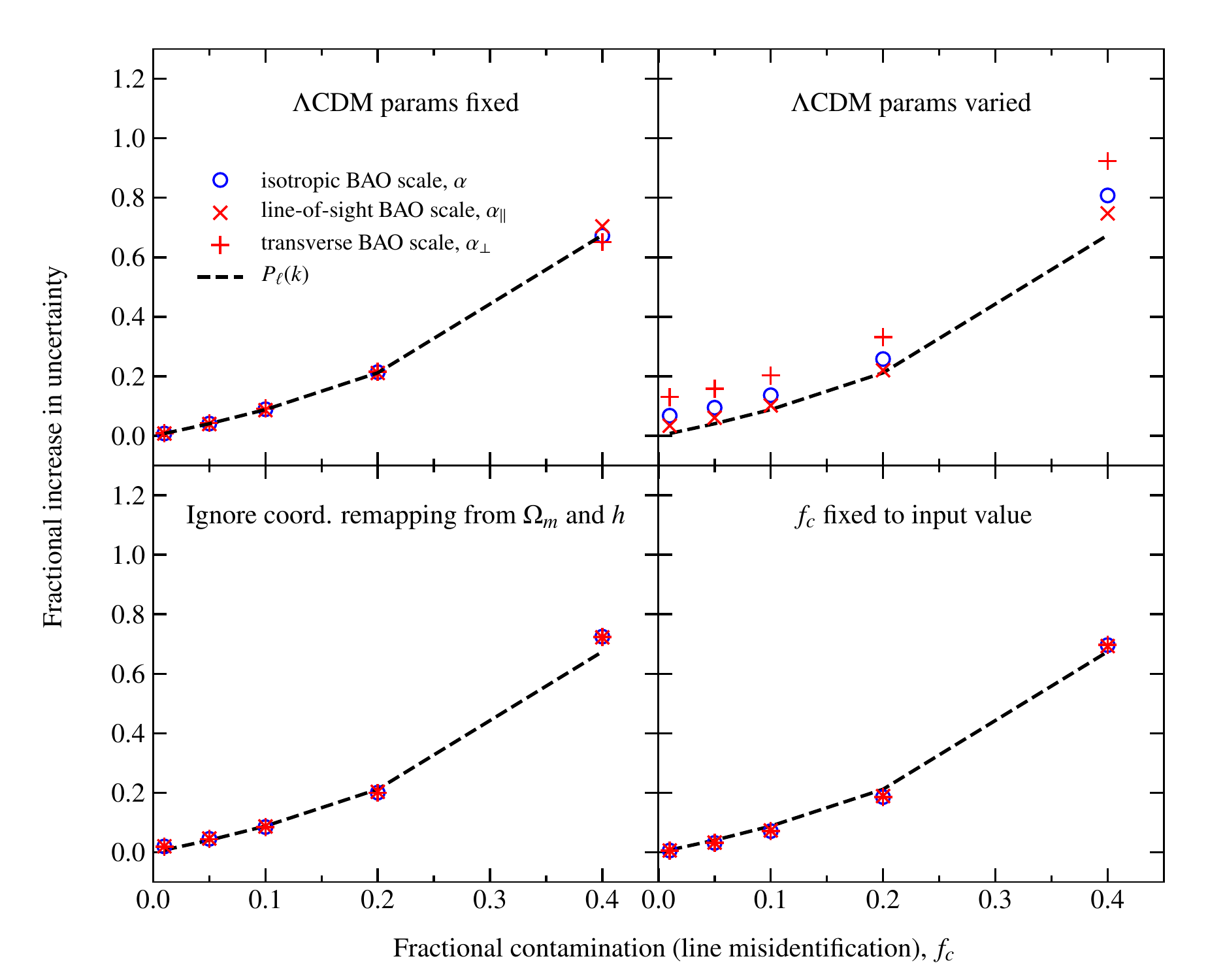}
\caption{Interlopers degrade constraints on the isotropic or anisotropic BAO scale, even if the interloper contribution is modeled correctly. Results are shown here for a Euclid-like [OIII] survey, contaminated by lower-redshift H$\alpha$ ELGs with fractional contamination values of 0.01, 0.05, 0.10, 0.20, and 0.40. The shape of the contamination power spectrum is taken to be known perfectly, multipoles $\ell=0,2,4$ are used and the bias of the [OIII] ELGs is marginalized over. \emph{Top left:} When the \lcdm\ parameters are held fixed, the degradation of the BAO scale constraints closely follows the loss of total signal in the target [OIII] multipole power spectra (black dashed line). \emph{Top right:} When \lcdm\ parameters are all marginalized over, with no external priors, the BAO scale recovery is further degraded from a degeneracy with $f_c$, even if the true $f_c$ is small. \emph{Bottom left:} If we ignore the angle-to-position ELG coordinate remapping effect from varying $\Omega_m$ and $h$ (i.e. the change $D_A$ and $H$ at the survey redshift), so that these parameters only affect the matter power spectrum shape, the degeneracy with $f_c$ disappears. \emph{Bottom right:} Fixing $f_c$ to the true value means BAO parameters are recovered as well as in the case where \lcdm\ parameters are fixed. Fixing $\Omega_b$ and $n_s$ produces results similar to this panel (see text).}
\end{figure*}

\subsection{Recovery of BAO scale}

Figure~2 shows the forecast increase in uncertainty in the BAO scale parameters $\alpha$, $\alpha_{\bot}$, and $\alpha_{\parallel}$, defined in Section~2.7, as a function of the input $f_c$, for the Euclid [OIII] survey. Note that $\alpha_{\bot}$ and $\alpha_{\parallel}$ are always varied together. The different panels show results for different assumptions about the broadband power spectrum, discussed below. In all cases we plot the ratio of the uncertainty obtained from inverting the Fisher matrix defined in equation (14) to the corresponding uncertainty in a forecast where the sample is pure [OIII] and $f_c$ is known to be zero. Here we assume perfect knowledge of the remapped interloper power spectrum. This assumption is discussed in more detail in Section~4.4, below.

We first considered an optimistic scenario in which the matter power spectrum is taken to be known perfectly (top left panel of Fig.~2). In this case, the parameters that are varied in the Fisher forecast are either $\alpha$ (isotropic) or $\alpha_{\bot}$ plus $\alpha_{\parallel}$ (anisotropic), as well as $b_g$, and $f_c$. The uncertainties in the BAO scale closely follow the loss of overall constraining power in the target [OIII] power spectrum (dashed black line, defined in equation 21).

Secondly, we considered a more pessimistic scenario in which all the \lcdm\ parameters ($\Omega_b$, $\Omega_m$, $h$, $n_s$, and $\sigma_8$) are marginalized over, which acts to modify the shape and amplitude of $P_m(k)$, as well as the contrast (sharpness) of the BAO wiggles. In existing surveys the power spectrum is often modeled using a fixed fiducial model spectrum multiplied by a low-order polynomial function and exponential \citep[e.g.,][]{anderson/etal:2014}. Marginalizing over the polynomial coefficients and exponential cut-off scale allows the BAO scale to be extracted while allowing for imperfect modeling of the broadband power spectrum, particularly non-linearities. We expect marginalizing over the parameters determining $P_m(k)$ (without any external constraints, for example from the CMB) to achieve approximately the same effect in our forecasts.

Clearly, the BAO scale constraints will be substantially degraded when the \lcdm\ parameters are marginalized over, since a phenomenological shift in the BAO scale can be largely compensated by shifts in the \lcdm\ parameters, especially for a limited range of $k$ values. In other words, the BAO scale is a large part of how the galaxy power spectrum constrains the \lcdm\ parameters. Again, here we are interested in how the BAO constraints are further degraded in the presence of interlopers, not the absolute precision of the BAO recovery.

Varying $\Omega_m$ or $h$ also changes how angular position and redshift transform into three-dimensional position, and thus leads to anisotropic rescaling of the whole target ELG power spectrum through changes to $D_M(z)$ and $D_H(z)$ relative to the fiducial model. Mathematically this effect is equivalent to the interloper remapping in Section~2.3. Since anisotropic information is important for identifying the presence of interlopers one might imagine that including the changes in $D_M(z)$ and $D_H(z)$ leads to a stronger degeneracy between $f_c$ and $\Omega_m$ or $h$. This is indeed the case, and a partial degeneracy between $\alpha$ and $f_c$ also opens up (top right panel of Fig.~2). The uncertainties in the BAO scale, particularly the transverse scale, are increased beyond the loss of information in the target power spectrum, even if the true contamination is small or zero.

To verify the importance of this coordinate remapping effect, we forecast constraints without including it, so that varying $\Omega_m$ and $h$ only impacts the shape of the matter power spectrum. The results are shown in the bottom left panel of Figure~2. While there is a large degeneracy among the \lcdm\ parameters determining the power spectrum shape, the addition of interlopers does not degrade the BAO constraints beyond the power spectrum signal-to-noise loss.

The bottom right panel of Figure~2 shows results where all the \lcdm\ params are varied but $f_c$ is held fixed to the true value (approximating the case where we have a precise external constraint on $f_c$). In this case, despite the freedom allowed in the shape of $P_m(k)$, and the anisotropic rescaling of the matter power spectrum with changes in $\Omega_m$ and $h$, discussed above, the BAO uncertainties increase with $f_c$ following the power spectrum uncertainties. This illustrates that degeneracy with $f_c$ is what causes the increased uncertainties in the top right panel.

We finally performed a forecast where $\Omega_b$ and $n_s$ are held fixed, so only $\Omega_m$, $h$, and $\sigma_8$ are varied. The results are not shown in Figure~2 but are virtually indistinguishable from the bottom right panel. Fixing $\Omega_b$ and $n_s$ substantially reduces degeneracies between parameters determining the shape of the ELG power spectrum. A change in the BAO scale cannot be compensated by a change in the other parameters in the way that is possible when all the \lcdm\ parameters are free. Fixing $\Omega_b$ and $n_s$ is motivated by the fact that the ELG power spectrum over a modest range of scales does not precisely constrain either parameter, while they are both determined extremely precisely by modern CMB power spectrum constraints. Additionally the CMB constraints on these parameters are fairly robust to modifications in the cosmological model, particularly low-redshift modifications such as evolution in dark energy density that the BAO surveys are aiming to probe \citep[see, e.g., Table~5 of][]{planck/6:2018}. Strictly speaking, the CMB spectra are sensitive to $\Omega_bh^2$ rather than $\Omega_b$, however the Fisher forecasts for the BAO scale are the same in either case.

In summary, if a perfect template for the interloper power is available, BAO constraints are not degraded beyond the loss of information in the power spectrum provided either an external constraint on $f_c$, or an external constraint on the broadband power spectrum shape (here we considered fixing $\Omega_b$ and $n_s$), are available. 

\subsection{Recovery of cosmological constraints from redshift-space distortions}

Figure~3 shows forecast constraints on the RSD parameter $f\sigma_8$ in a similar way to Figure~2, comparing uncertainties against a pure [OIII] ELG survey with $f_c$ fixed to zero. Note that we are not jointly varying the BAO scale parameters $\alpha$, $\alpha_{\bot}$, or $\alpha_{\parallel}$ in these fits. Also, fractional errors on $f\sigma_8(z)$ and $\sigma_8(z=0)$ are equal in our forecasts, since these quantities are related by a factor that is a function of $\Omega_m$ only (the combination of $f$ and the linear growth factor that determines the redshift-dependence of $\sigma_8$). Results are shown for $f_c=0.15$ in addition to the values used in Figure~2.

When the \lcdm\ parameters are all varied the $f\sigma_8$ constraints are completely degraded for low values of $f_c$, and somewhat exceed the power spectrum constraints for larger values (red points and line in Fig.~3). When the true value of $f_c$ is small, the presence of interlopers cannot be detected at high significance from the contaminated power spectra alone. Consequently there is close to a complete degeneracy between the amplitude of the matter power spectrum (here parameterized by $f\sigma_8$) and $(1-f_c)^2$, which multiplies the target ELG power in the first term of equation (8). When the true value of $f_c$ is larger, the interlopers are detected, however a partial degeneracy with the power spectrum amplitude remains.

The blue circles in Figure~3 shows that, when $f_c$ is fixed, the uncertainty in $f\sigma_8$ follows the power spectrum uncertainty. Since $f\sigma_8$ constraints depend on anisotropic information, in our case from the relative amplitudes of the different multipoles of the power spectrum, one would expect introducing additional anisotropic contributions from interlopers to impact constraints more strongly than in the BAO case. We have shown that this is indeed the case. \emph{The fact that the degeneracy gets worse for lower values of $f_c$ strongly motivates exploring external constraints on $f_c$}.

We also considered a range of alternatives, not shown in Figure~3, including: (i) fixing $\Omega_b$ and $n_s$, (ii) fixing $b_g$, and (iii) removing the anisotropic rescaling effect of $\Omega_m$ and $h$, that is, assuming the conversion between angular and three-dimensional coordinates is known perfectly. None of these changes remove the strong degeneracy between $f_c$ and $f\sigma_8$ when $f_c$ is small. Fixing $b_g$, for example, does improve precision of $f\sigma_8$ recovery, however it does so in roughly the same way in both the pure [OIII] and contaminated cases, and so does not bring the red points and line in Figure~3 into agreement with the dashed black line.

The Fisher forecasts do not account for the fact that $f_c$ must be positive, and may also be unreliable when there is a severe degeneracy since the response of the data to a small change in parameter values is assumed to be linear. We therefore wanted an independent verification of the behavior shown for low $f_c$ in Figure~3. We examined the relationship between $f\sigma_8$ and $f_c$ by drawing 1000 Gaussian realizations of $P_{t,\ell}(k_b)$ from the covariance matrix, $\mathcal{C}_{\ell\ell'}(k_b,k_b)$, and numerically finding the maximum-likelihood values of $f\sigma_8$ and $f_c$ in each case, taking all other parameters as fixed, and requiring $f_c\geq0$. We tested the case where $f_c=0.2$ and found good agreement between the uncertainty in $f\sigma_8$ obtained from the Fisher forecast and the spread in the maximum-likelihood values from the 1000 realizations. We then tested the case where there is no contamination, $f_c=0$, but $f_c$ is still varied, generating 1000 pure [OIII] realizations. In around half the realizations, the best-fit $f_c$ was essentially zero, and the spread in values of $f\sigma_8$ was consistent with the case where only $f\sigma_8$ was varied. In the other half of the realizations, however, the best-fit value of $f_c$ was non-zero, and $f\sigma_8$ was biased high to compensate. For these realizations, the spread in $f\sigma_8$ values was over ten times the spread with $f_c$ held fixed to zero, confirming the degeneracy indicated by the Fisher calculations.

\begin{figure}
\includegraphics{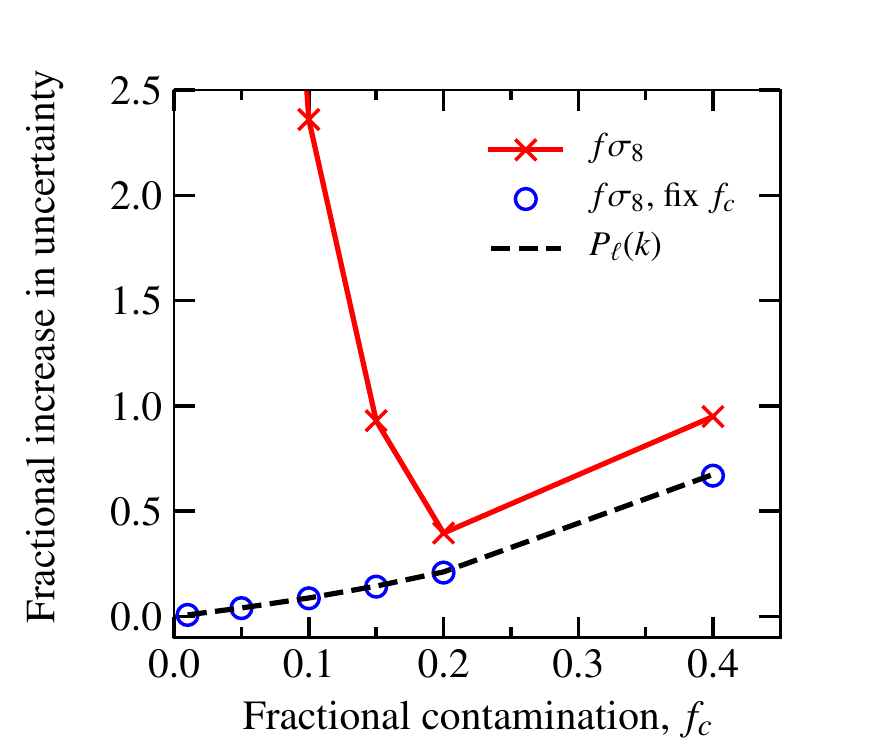}
\caption{Fisher forecasts of RSD parameter $f\sigma_8$ from a Euclid-like [OIII] survey, as a function of fractional contamination by H$\alpha$ ELGs. The dashed black line shows the loss of signal-to-noise ratio in the power spectrum, as in Figure~2. The red line shows the increase in $f\sigma_8$ uncertainty compared to a pure [OIII] survey with $f_c=0$ when all the other \lcdm\ parameters, namely $\Omega_m$, $\Omega_b$, $h$, and $n_s$, are varied along with $b_g$ and $f_c$. There is a near-complete degeneracy between $f\sigma_8$ and $f_c$ when the true $f_c$ is small, discussed further in the text. The blue circles show that when $f_c$ is fixed to the input value the $f\sigma_8$ constraints follow the power spectrum constraints.}
\end{figure}

One might ask whether the bias on $\sigma_8$ or $f\sigma_8$ from ignoring the contamination in the fitting might be acceptable when $f_c$ is small. Taking partial derivatives of the power spectrum, the bias in $\sigma_8$ from ignoring the contamination when $f_c$ is small is given by $\delta \sigma_8/\sigma_8=-f_c+\mathcal{O}(f_c^2)$. Note that there is no dependence on the interloper power to first order in $f_c$ (equation 8). Our Fisher forecast predicts a $1\sigma$ statistical uncertainty of $\delta\sigma_8/\sigma_8=0.021$ for the Euclid [OIII] survey in the case where other \lcdm\ parameters are marginalized over. While this forecast is optimistic (Section~2.1), it implies that a significant bias could result from ignoring interlopers if $f_c$ is not known to be smaller than a fraction of a percent. This is consistent with the calculations of P16, who found that even a $0.15-0.30$\% interloper fraction could bias the growth rate measurements by more than 10\% of the statistical error.

\subsection{Constraining the interloper power spectrum and $f_c$ in cross-correlation}

In the calculations above we fixed the interloper power spectrum, motivated in part by the fact for the Euclid or WFIRST [OIII] surveys, which are among the most prone to redshift misidentification (e.g., P16), the primary contaminant is H$\alpha$ ELGs, which are the main cosmology target of these experiments. As a result, measurements of the H$\alpha$ power spectrum will be obtained at a higher precision than the contaminated [OIII].

If a separate catalog of the ELGs that were misidentified is available, a cross-correlation with the contaminated catalog can also provide a precise constraint on $f_c$. Consider taking the Euclid H$\alpha$ catalog from the same patch of sky as the [OIII] catalog and remapping the three-dimensional coordinates `by-hand' as if the ELGs in the main H$\alpha$ sample had all been misidentified as [OIII]. Assuming the target and interloper populations do not overlap in redshift, the expected cross-power spectrum between this remapped H$\alpha$ sample and the contaminated [OIII] sample will be given by
\be
\left\langle \hat{P}_{\times,\ell}(k)\right\rangle=f_c\gamma_{\bot}^2\gamma_{\parallel}P_{\rm int,\ell}(\gamma_{\bot}{\bf k}_{\bot},\gamma_{\parallel}{\bf k}_{\parallel}),
\ee
that is, like the second term of equation (8), except scaling like $f_c$ rather than $f_c^2$. The auto-power spectrum of the remapped H$\alpha$ catalog, on the other hand, does not depend on $f_c$:
\be
\left\langle \hat{P}_{\rm auto,\ell}(k)\right\rangle=\gamma_{\bot}^2\gamma_{\parallel}P_{\rm int,\ell}(\gamma_{\bot}{\bf k}_{\bot},\gamma_{\parallel}{\bf k}_{\parallel}).
\ee

The combination of the contaminated power spectra, $\hat{P}_{t,\ell}$, remapped-contaminated cross-spectrum, $\hat{P}_{\times,\ell}$, and remapped auto-spectrum, $\hat{P}_{\rm auto,\ell}$, can simultaneously constrain the shape of the interloper power, $f_c$, and the parameters we are trying to measure from the contaminated sample. We demonstrate this for constraints on \lcdm\ parameters from a Euclid-like [OIII] survey with $f_c=0.2$ in Figure~4. Results are similar for BAO parameters. Here we do not assume a particular model for the H$\alpha$ power spectrum and instead consider the power spectrum in each bin of the monopole, quadrupole, and hexadecapole power spectrum as a free parameter. Even in this more conservative case the addition of the $\hat{P}_{\times,\ell}$ and $\hat{P}_{\rm auto,\ell}$ spectra allow cosmological constraints to be recovered from the contaminated [OIII] sample as if $f_c$ was known perfectly. In fact, there is a small (percent level) improvement in constraints over the forecast for the contaminated [OIII] sample with $f_c$ fixed. In other words, we can do a little better than the black lines in Figures~2 and 3 would suggest. This is because adding precise measurements of the H$\alpha$ power from galaxies tracing the same modes of the density field as the misidentified H$\alpha$ ELGs also effectively removes the H$\alpha$ sample variance as an error source in the contaminated power spectra. Unfortunately, the improvement is small because the H$\alpha$ interloper sample variance is only a small contribution to the total power spectrum error budget.

One possible concern with this approach is uncertainty in the coordinate remapping that we should apply by-hand to the separate catalog of H$\alpha$ ELGs (the $\gamma_{\bot}$ and $\gamma_{\parallel}$ factors). We cannot completely ignore this uncertainty since, if the remapping was known perfectly in advance, there would be little motivation to make the BAO measurement from secondary samples like the Euclid [OIII] in the first place. Fortunately, the remapping can be determined empirically simply by varying the transverse and line-of-sight rescaling and recomputing the remapped-contaminated cross-spectrum. If either of the rescaling factors is substantially incorrect, modes of the remapped ELG overdensity field will not fall in the same power spectrum bin as the corresponding modes of the interlopers in the contaminated sample, and the cross-spectrum will be consistent with zero. Since coarse binning of the power spectrum is already undesirable for BAO constraints, we do not expect the remapping uncertainty to be a limitation, although a detailed demonstration of this is left as a future project.

\begin{figure}
\includegraphics{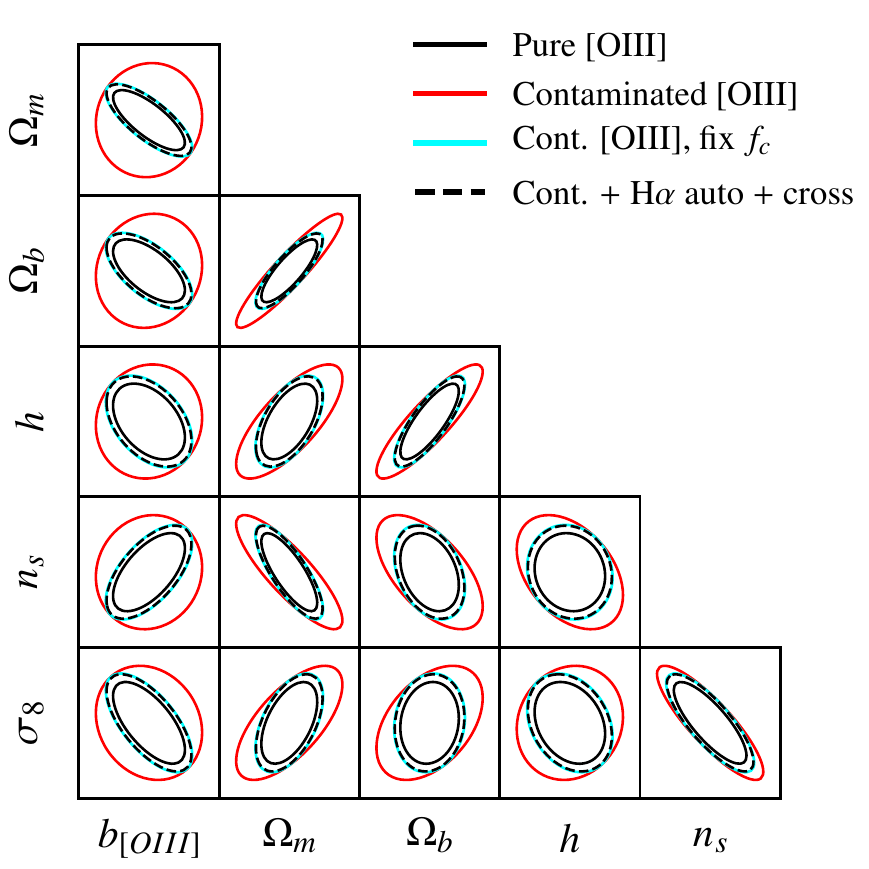}
\caption{External catalog of the misidentified ELGs helps break degeneracies between $f_c$ and cosmological parameters. Adding the cross-spectrum between the contaminated and external catalog, and the auto-spectrum of the external catalog (equations 22 and 23) constrains parameters slightly better than fixing $f_c$ to the input value, even though the bandpowers of the contaminating power spectrum are also marginalized over in this case. These Fisher forecasts were performed for a Euclid-like [OIII] survey with contamination fraction $f_c=0.2$ and an external pure H$\alpha$ catalog.}
\end{figure}

\subsection{Effect of changing range of $k$ values}

We investigated the effect of changing the range of scales used in the forecasting. We considered lowering $k_{\rm max}$, representing excluding information from small scales that are challenging to model due to nonlinear clustering. We also considered increasing $k_{\rm min}$, which has less physical motivation and was done primarily to test assumptions discussed in Section~2.4. The main conclusions regarding the BAO and RSD constraints discussed earlier in this section still held and no new behavior was observed. Note that we are always comparing against pure [OIII] constraints with the same $k_{\rm min}$ and $k_{\rm max}$.

Quantitatively, the main effect we found from changing the range of scales was a worsening of the degeneracy between $\alpha_{\bot}$ and $f_c$ in the anisotropic BAO case where \lcdm\ parameters are varied. This is not surprising given we are removing information by restricting the available $k$ modes. We give some example results here for $f_c=0.2$. For the $k_{\rm max}=0.3$~$h$Mpc$^{-1}$ case shown in the top right panel of Figure~2, the $\alpha_{\bot}$ error is increased beyond the loss of information in the power spectrum by 10\%. For $k_{\rm max}=0.225$~$h$Mpc$^{-1}$, the corresponding increase is 19\%. For $k_{\rm max}=0.15$~$h$Mpc$^{-1}$, the $\alpha_{\bot}$ error is increased by 46\%. Increasing $k_{\rm min}$ from 0.005 up to 0.1~$h$Mpc$^{-1}$ (holding $k_{\rm max}$ fixed to 0.3~$h$Mpc$^{-1}$) similarly resulted in an increase of 46\%. For $\alpha_{\parallel}$, or $\alpha$ in the isotropic BAO case, the results are consistent with Figure~2 within a few percent. Furthermore, the increase in BAO scale (isotropic or anisotropic) for the case with $\Omega_b$ and $n_s$ fixed, or with $f_c$ fixed, are consistent with the increase in power spectrum uncertainties within a few percent, also similar to Figure~2, even for $k_{\rm max}=0.15$~$h$Mpc$^{-1}$. External constraints on portions of the \lcdm\ parameter space or $f_c$ are effective at breaking the $f_c-\alpha_{\bot}$ degeneracy even when the range of scales is limited.

We found that reducing $k_{\rm max}$ did not significantly impact the degeneracy between $f_c$ and $f\sigma_8$ shown in Figure~3. A more detailed analysis of the effect of including nonlinearity in the ELG power spectrum, and marginalization over parameters that describe how the nonlinearity is modeled, particularly for constraints on neutrino mass, are left to future analysis including survey-specific systematic effects.

\section{Discussion}

In Section~4 we performed calculations for a Euclid-like [OIII] ELG survey ($1.5<z<2.3$) contaminated by H$\alpha$ ($0.9<z<1.5$) and identified two ways interlopers degrade cosmological constraints, even when included in the likelihood modeling: loss of signal-to-noise in the target power spectrum, and additional degeneracy with the contamination fraction, $f_c$. In this section we expand our analysis to consider implications for other lines and surveys.

\subsection{Varying galaxy density and implications for WFIRST}

We repeated calculations from Section~4 with higher [OIII] source densities, up to ten times the original 282~deg$^{-2}$. This would roughly correspond to a flux limit of $1.0\times10^{-16}$~erg~s$^{-1}$~cm$^{-2}$ based on Table~2 of \cite{colbert/etal:2013}, which predicted 3056~deg$^{-2}$. While the absolute constraining power of the survey would dramatically increase, the effect of interlopers for a given $f_c$ is very similar to the results shown in Figures~2 and 3. Compared to the increase in power spectrum errors, the BAO errors for the case where \lcdm\ parameters are all varied are slightly increased with the higher galaxy density. For example the increase in $\alpha_{\bot}$ error is around 20\% larger than the increase in power spectrum error, whereas in Figure~2 the difference is more like 10-15\% (compare red cross to black line). The degeneracy between $f_c$ and power spectrum amplitude for the RSD constraint is present in the same way.

We have not repeated calculations for the WFIRST High Latitude Survey volume and expected galaxy density (Section~3.3), but since the number density of [OIII] ELGs is around twice the density we used in Section~4 we do not expect any substantially different behavior for a given $f_c$. That said, the higher signal-to-noise cut for ELG detection in WFIRST would both reduce the probability of misidentification occurring, and make dealing with additional complications, such as spatial variation in $f_c$, less challenging.

\subsection{Impact for different target lines}

\begin{figure*}
\hspace{-0.6cm}
\includegraphics{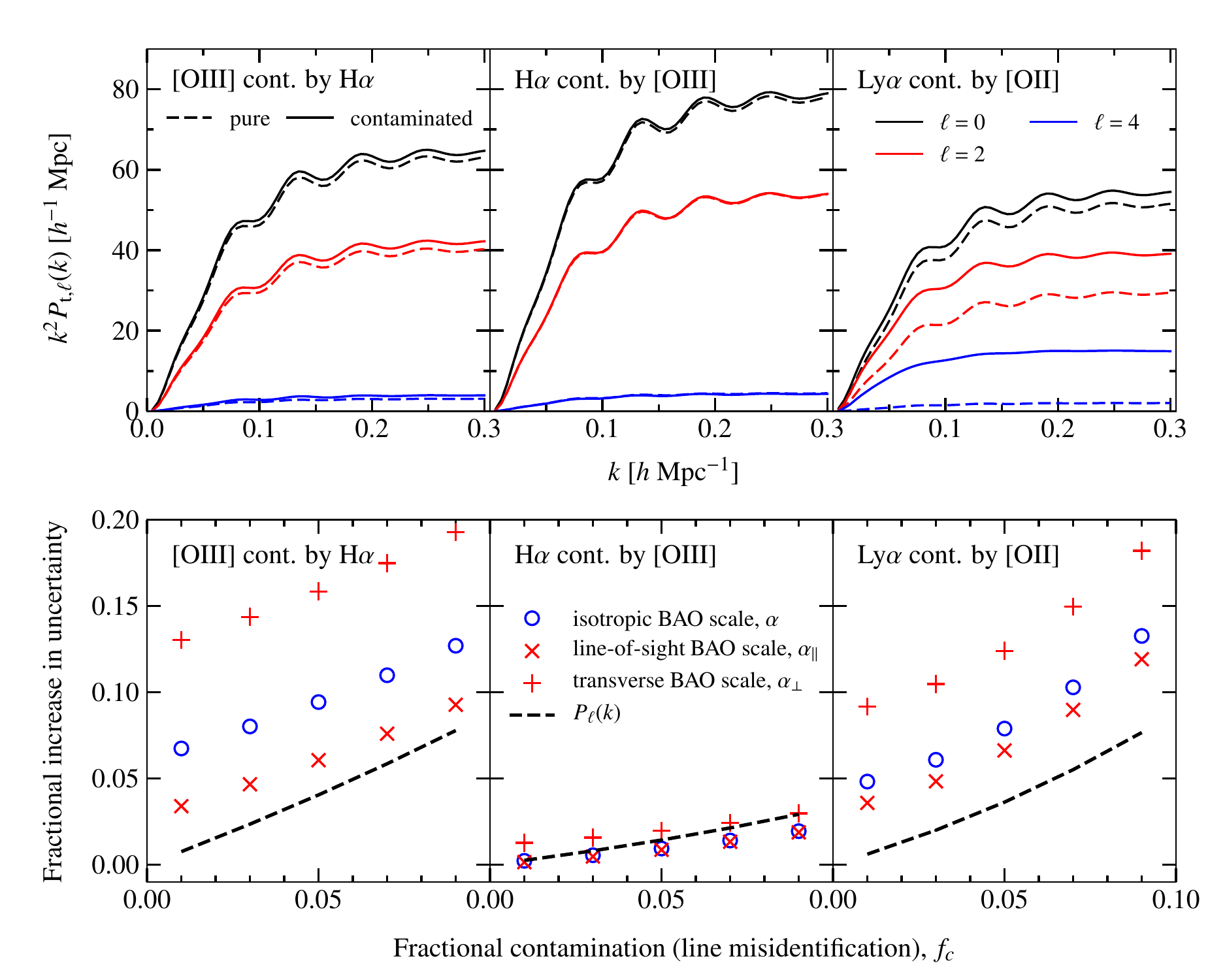}
\caption{\emph{Top row:} Comparison of pure (dashed lines) and contaminated (solid lines) multipole power spectra for (left to right) Euclid-like [OIII] sample ($z=1.9$) contaminated by H$\alpha$ ($z=1.2$), Euclid-like H$\alpha$ sample contaminated by [OIII], and and HETDEX-like Ly$\alpha$ sample ($z=2.7$) contaminated by [OII] ($z=0.2$). A contamination fraction $f_c=0.12$ is shown in each case. The contaminated power spectra has been divided by the overall suppression factor $(1-f_c)^2$ (see equation 8) to make the impact of interlopers on different multipoles clearer. \emph{Bottom row:} Increase in BAO scale errors for $f_c=0.01,0.03,0.05,0.07,0.09$ when \lcdm\ parameters and $f_c$ are marginalized over (compare to top right panel of Fig.~2). Loss of signal-to-noise ratio in the target ELG power spectra is shown for comparison (dashed lines, see equation 21).}
\end{figure*}

The top row of panels of Figure~5 shows the effect of interlopers on the multipole power spectra for three different ELG samples: a Euclid-like [OIII] sample contaminated by H$\alpha$ (as shown in Section~4), a Euclid-like H$\alpha$ sample contaminated by [OIII], and a HETDEX-like Ly$\alpha$ sample contaminated by [OII]. We adopted a contamination fraction of $f_c=0.12$ so that the difference between the pure and contaminated samples is more apparent by eye, recognizing that this value is unrealistically large for HETDEX, where contamination is only expected at the percent level \citep{leung/etal:2017}. Additionally, we note that a contamination fraction $f_c\simeq0.07$ for the Euclid-like H$\alpha$ sample would require every single [OIII] ELG to be misidentified as H$\alpha$ for the number densities shown in Table~1. Investigating larger contamination fractions is motivated by the substantial uncertainty in the source count forecasts. Since the main effect of low-level contamination is a suppression of the target ELG power by $(1-f_c)^2$ (equation 8), we divided the contaminated spectra by this factor, so that the difference between the pure and contaminated spectra comes solely from the interloper contribution (second term of equation 8).

The effects of interlopers on the power spectrum for a given $f_c$ depends on the redshift of the target and interloper line primarily in two ways. Firstly, the volume factor $\gamma_{\bot}^2\gamma_{\parallel}$ from remapping interloper coordinates to the target line redshift rescales the whole interloper power spectrum at each multipole in equation (8). When [OIII] is viewed as a contaminant to H$\alpha$ (middle panel), rather than the other way around (left panel and Section~4), the remapping parameters $\gamma_{\bot}$ and $\gamma_{\parallel}$ are inverted. The volume factor is decreased from $1.7$ to $1/1.7\simeq0.6$, a difference of nearly a factor of three. Secondly, when the interloper ELGs are at higher redshift than the targets (e.g., when [OIII] contaminates H$\alpha$), the quadrupole and hexadecapole power contributed by the interlopers is suppressed relative to the monopole, and may become negative. This arises because $\gamma_{\bot}<1$, while (as in all cases considered) $\gamma_{\parallel}\simeq1$. Transverse modes receive an enhancement because the matter power spectrum $P_m(k)$ is falling with $k$ for $k\geq0.02$~$h$Mpc$^{-1}$ (past the peak of the power spectrum, note that we show $k^2P(k)$ in plots, not $P(k)$ itself). Apart from the first few $k$ bins, then, the power at $\gamma_{\bot}k$ is larger than the power at $k$. This effect counteracts the usual RSD enhancement of line-of-sight power, which is what produces positive $\ell=2$ and $\ell=4$ power for analysis in the correct coordinate system, and is why the $\ell=2$ and $\ell=4$ spectra are largely unchanged by interlopers in the middle panel.

The converse happens for interlopers at lower redshift than the targets, and the power at $\ell=2$ and $\ell=4$ is enhanced relative to $\ell=0$ as a result. This is clearly apparent in the HETDEX case because of the large difference in redshift between the targets ($1.9<z<3.5$) and interlopers ($z<0.5$). The remapping factors are $\gamma_{\bot}=7.3$ and $\gamma_{\parallel}=0.9$, so that modes at $\gamma_{\bot}k$ have much lower power than $\gamma_{\parallel}k$.

The panels in the bottom row of Figure~5 show the forecast increase in BAO scale constraints as a function of contamination fraction, $f_c$, for the case where \lcdm\ parameters are marginalized over in addition to $f_c$ itself. The results for Euclid H$\alpha$ contaminating [OIII] are the same as in the top right panel of Figure~2. To highlight the impact of interlopers we show fractional increases in BAO error forecasts over pure samples for each target line, as in Section~4. We include the fractional loss of overall power spectrum signal-to-noise ratio for comparison (equation 21, dashed black lines).

A given $f_c$ leads to a far smaller increase in BAO scale uncertainty for the case where [OIII] is contaminating H$\alpha$, compared to when the H$\alpha$ is contaminating [OIII]. The degeneracy between $f_c$ and the \lcdm\ parameters, particularly $\Omega_m$ and $h$, is weaker. In other words, the increase in the monopole power caused by the higher-redshift [OIII] interlopers in the middle panel of the top row of Figure~5 is harder to mimic through rearranging other parameters. We note that the number density of H$\alpha$ ELGs in our Euclid-like forecasts is almost 14 times higher than [OIII] (3900~deg$^{-2}$ compared to 282~deg$^{-2}$, Table~1), which means the shot-noise errors on the power spectrum bins are far lower. We performed calculations with the H$\alpha$ density decreased by hand to match [OIII] to understand how this number density difference impacts the BAO error forecasting. Even with this change, the increase in BAO uncertainty for the [OIII] targets was worse than for the H$\alpha$ targets by a factor of two to four for a given $f_c$, verifying the importance of the relationship between target and interloper redshift, and the $\gamma_{\bot}$ and $\gamma_{\parallel}$ factors, in determining the degeneracy between BAO $\alpha$ factors and $f_c$. As in Section~5.1, we emphasize here that looking at the ratio of forecast uncertainties for contaminated sample to pure sample substantially reduces the importance of effects like number density that impact both.

The forecasts for HETDEX BAO uncertainties are fairly similar to the Euclid [OIII] targets despite the large difference in volume factor $\gamma_{\bot}^2\gamma_{\parallel}$ mentioned above. While a given $f_c$ produces a larger difference in the multipole power spectra for HETDEX, the effect on the power spectrum errors is fairly small in either case for $f_c\lesssim0.1$. We experimented with varying the HETDEX and [OIII] ELG number densities and found that this did not have a large impact on the results shown in Figure~5. We note that the impact of HETDEX interlopers for a given $f_c$ may be inaccurate here for two reasons. Firstly, our use of the linear matter power spectrum in calculations is a poor approximation for the remapped [OII] ELGs because the large transverse dilation factor $\gamma_{\bot}\simeq7$ means smaller, more highly nonlinear, scales are being probed for a given $k$ in the coordinates of the Ly$\alpha$ targets. Additionally, the value of $\gamma_{\bot}$ itself is highly sensitive to the effective redshift adopted for the [OII] sample since the sample extends to redshift zero, for example \cite{leung/etal:2017} found $\gamma_{\bot}$ between 5.3 and 29.8 for [OII] galaxies at redshift 0.305 and 0.044, respectively.

Fixing $f_c$ brings the increase in BAO scale uncertainty into agreement with the black dashed lines for the points shown in the left and right panels of the bottom row of Figure~5, as in Figure~2. There is little effect for the points shown in the middle panel because the degeneracy with $f_c$ is weaker. In fact, one can see in this panel that some of the BAO points lie below the black dashed line even when $f_c$ is varied. Results for RSD are not shown in Figure~5 but the degeneracy between $f_c$ and $f\sigma_8$ is present and degrades constraints by a factor of at least a few for $f_c<0.1$, as in Figure~3. \emph{While the details of the interloper remapping, and absolute constraining power of the surveys, vary significantly, an external constraint on $f_c$ is essential for recovering the RSD information in all cases we considered.}

\subsection{Future directions}

We have focussed on the BAO scale and RSD as the primary cosmological observables from ELG surveys, and performed forecasts for the impact of line misidentification with some simplifying assumptions. In the future it would be useful to extend the investigation to other aspects of galaxy clustering, including: BAO reconstruction \citep[e.g.,][]{padmanabhan/etal:2012}, constraints from the bispectrum and other higher-order statistics, choice of redshift binning, and clustering on scales larger than the BAO \citep[including associated systematics, e.g.,][]{kalus/etal:2019}. 

We have highlighted the importance of an external constraint on $f_c$ for RSD constraints. In Section~4.4 we described how a cross-correlation with a separate catalog containing the misidentified line can achieve this, however such a catalog may not be available for low-level contaminants. Another approach is to use deeper sub-surveys, where multiple lines are detected and unambiguously identified in each ELG spectrum. An estimate of the expected contamination level for the main survey can then be made for each potential interloper line, for example by adding artificial noise to the deeper spectra. The Euclid survey strategy includes observing several small regions (tens of square degrees) two magnitudes deeper than the main survey \citep{laureijs/etal:2011}. Demonstrating that low-level contaminants can be constrained sufficiently well to break the degeneracy between $f_c$ and $f\sigma_8$ using realistic simulated spectra would be valuable.

Simulated data are also required to understand the impact of spatial variation in misidentification rate, arising from the complex weighting or masking applied to real spectroscopic surveys, for example to deal with contamination from bright stars, Zodiacal dust emission, or variability in optical performance across the field of view of the instrument. One approach would be to extend galaxy weighting schemes, already adopted in current BAO surveys like BOSS \citep[e.g.,][]{ross/etal:2012}, in order to downweight ELGs with spectra more prone to misidentification (e.g., with lower signal-to-noise ratio) before computing clustering statistics. In principle this method could help recover some of the information lost in the presence of interlopers, although again this should be demonstrated quantitatively.

\section{Conclusions}

We used Fisher forecasts to investigate the impact of contamination of ELG catalogs due to spectroscopic line misidentification. We used redshift-space multipole power spectra to describe the anisotropic distortion of the interloper power spectrum due to incorrect mapping of galaxy angular position and redshift to three-dimensional position. Using calculations performed for a Euclid-like [OIII] survey contaminated by H$\alpha$ interlopers as an example, we found that cosmological constraints on the BAO scale and RSD parameter $f\sigma_8(z)$ are degraded in the presence of interlopers in two ways:
\begin{enumerate}[label=(\roman*)]
\item The presence of interlopers decreases the signal-to-noise ratio of the target ELG power spectra, even if $f_c$ and the shape of the interloper power are known perfectly. This is because the contaminated sample contains two populations tracing LSS at different redshifts, which do not correlate (neglecting small corrections, e.g., from gravitational lensing). Recovering this information requires additional information or weighting on a spectrum-by-spectrum basis. This will be investigated further in future work.
\item Degeneracy with $f_c$ can further degrade constraints. This increases BAO errors at the 10-20\% level for the case where \lcdm\ parameters determining the broadband power spectrum shape are marginalized over, although this assumes there is an external template for the interloper power. For the RSD, there is a near-complete degeneracy between $f_c$ and the power spectrum amplitude ($\sigma_8$ or $f\sigma_8$) when $f_c$ is too small for the presence of interlopers to be detected at high significance from the contaminated power spectrum alone.
\end{enumerate}

For the BAO, we found that external constraints on a portion of the \lcdm\ parameter space (e.g., $\Omega_b$ and $n_s$ from CMB measurements), or on $f_c$, remove the degeneracy with $f_c$. In the RSD case, only a constraint on $f_c$ achieved this, indicating that an estimate for the contamination fraction is important even for low-level contaminants. We considered constraining $f_c$ using cross-correlation with a separate catalog containing the misidentified ELGs. This appears to be an effective approach for constraining both $f_c$ and the shape of the interloper power for the Euclid and WFIRST H$\alpha$ and [OIII] ELGs, where both emission lines are cosmology targets, but may be challenging for minor contaminants.

More realistic calculations, including systematic effects that may impact line identification, such as spatial variability in spectra quality and signal-to-noise ratio, are necessary to quantify the impact of line misidentification more accurately. The formalism and results we have presented will help guide these future efforts.

The analysis in this paper complements that recently presented by \cite{grasshorn/etal:2019}, which focuses on calculations in two-dimensional Fourier space, starting from equation (6). Our results focus on BAO and RSD forecasts. \cite{grasshorn/etal:2019} examine the cross-correlation method discussed in Section~4.4 and address the case of two-way contamination for HETDEX and WFIRST. Note that the remapping factors we call $\gamma_{\bot}$ and $\gamma_{\parallel}$ are denoted $\alpha$ and $\beta$ by \cite{grasshorn/etal:2019}. The two projects arose from some earlier common discussion but the calculations were performed and manuscripts prepared independently.\\

This work was partly supported by NASA grants NNX14AB76A and NNX15AI57G. Addison and Bennett also acknowledge support from NASA ROSES grant 12-EUCLID12-0004. We would like to thank the referee for a careful reading of the manuscript and useful suggestions.\\
\\
\bibliographystyle{apj}

\appendix
\section{Power spectrum formalism}

The redshift dependence of quantities below is omitted for brevity. The fractional matter overdensity, $\delta_m({\bf x})$, is defined as
\be
\delta_m({\bf x})=\frac{\rho_m({\bf x})-\bar{\rho}_m}{\bar{\rho}_m},
\ee
where $\rho_m({\bf x})$ is the matter density at position ${\bf x}$ and $\bar{\rho}_m$ is the mean density (averaged over position). The Fourier coefficients of the matter overdensity field are denoted $\delta_m({\bf k})$. For statistically isotropic density fluctuations the matter power spectrum $P_m(k)$ is defined as
\be
\left\langle\delta_m({\bf k})\delta^*_m({\bf k'})\right\rangle=\delta^3_D({\bf k}-{\bf k'})P_m(k),
\ee
where the angle brackets represent averaging over a large number of realizations of the density field, $\delta^3_D$ is the three-dimensional Dirac delta, and $k=|{\bf k}|$.

Galaxies are biased tracers of the underlying matter density. In this work we assume galaxies are linear tracers, meaning that the galaxy and matter overdensities are related by a single scale-independent bias factor, $b_g$. For our purposes galaxies are discrete objects and the two-point product of galaxy overdensities produces a shot-noise component $1/n_g$, where $n_g$ is the galaxy number density:
\be
\left\langle\delta_g({\bf k})\delta^*_g({\bf k'})\right\rangle=\delta^3_D({\bf k}-{\bf k'})\left[P_g({\bf k}) + \frac{1}{n_g}\right].
\ee
This shot noise contribution is subtracted off whenever we are estimating the galaxy power spectrum, however it still contributes to the power spectrum uncertainties. We neglect any difference between the actual number density of galaxies in some surveyed volume and the mean number density in calculations.

Galaxy clustering is anisotropic due to peculiar motion of galaxies relative to the Hubble flow. Consequently $P_g({\bf k})$ depends on the direction of $\bf k$ rather than just the magnitude. In linear theory, the three-dimensional power spectrum is given by \citep[e.g.,][]{kaiser:1987,hamilton:1998}
\be
P_g({\bf k})=(1+\beta\mu_{\bf k}^2)^2P_g(k),
\ee
where $\beta=f/b_g$, with $f$ the derivative of the cosmological growth rate, and $\mu_{\bf k}$ is the cosine of the angle between the line of sight and $\bf k$ (i.e., $\mu_{\bf k}=k_z/\left|{\bf k}\right|$). To account for this anisotropy we work with the multipole galaxy power spectra, $P_{g,\ell}(k)$, which are related to the full three-dimensional power spectrum, $P({\bf k})$, by
\be
P_{g,\ell}(k)=\frac{2\ell+1}{2}\int_{-1}^1d\mu_{\bf k}P_g({\bf k})\mathcal{P}_{\ell}(\mu_{\bf k}),
\ee
where $\mathcal{P}_{\ell}$ are Legendre polynomials. In linear theory only the monopole, quadrupole, and hexadecapole moments ($\ell=0,2,4$) of the multipole power spectra are non-zero. The integrals in (A5) can be evaluated analytically and produce
\be
\begin{split}
P_{g,0}(k)&=\left(1+\frac{2}{3}\beta+\frac{1}{5}\beta^2\right)b_g^2P_m(k)\\
P_{g,2}(k)&=\left(\frac{4}{3}\beta+\frac{4}{7}\beta^2\right)b_g^2P_m(k)\\
P_{g,4}(k)&=\frac{8}{35}\beta^2b_g^2P_m(k).\\
\end{split}
\ee
To estimate the multipole power spectra given a particular realization of the galaxy overdensity field $\delta_{g,{\bf k}}$ we define an estimator
\be
\hat{P}_{g,\ell}(k)=\frac{2\ell+1}{2}\int_{-1}^1d\mu_{\bf k}\,\delta_g({\bf k})\delta^*_g({\bf k})\mathcal{P}_{\ell}(\mu_{\bf k})-\frac{\delta_{\ell0}}{n_g}.
\ee
Thanks to the shot-noise subtraction this results in an unbiased estimate, so that $\left\langle\hat{P}_{g,\ell}(k)\right\rangle=P_{g,\ell}(k)$. We can then compute the covariance of the estimator as 
\be
\begin{split}
\mathcal{C}_{\ell,\ell'}(k,k')&=\left\langle\left(\hat{P}_{g,\ell}(k)-\left\langle\hat{P}_{g,\ell}(k)\right\rangle\right)\left(\hat{P}_{g,\ell'}(k')-\left\langle\hat{P}_{g,\ell'}(k')\right\rangle\right)\right\rangle\\
&=\frac{(2\ell+1)(2\ell'+1)}{4}\int_{-1}^1\int_{-1}^1d\mu_{\bf k}\, d\mu_{\bf k'}\left(\left\langle\delta_g({\bf k})\delta^*_g({\bf k})\delta_g({\bf k'})\delta^*_g({\bf k'})\right\rangle-\left\langle\delta_g({\bf k})\delta^*_g({\bf k})\right\rangle\left\langle\delta_g({\bf k'})\delta^*_g({\bf k'})\right\rangle\right)\mathcal{P}_{\ell}(\mu_{\bf k})\mathcal{P}_{\ell'}(\mu_{\bf k'})\\
&=\frac{(2\ell+1)(2\ell'+1)}{4}\int_{-1}^1\int_{-1}^1d\mu_{\bf k}\, d\mu_{\bf k'}\,\left[\delta^3_D({\bf k}-{\bf k'})+\delta^3_D({\bf k}+{\bf k'})\right]\left[(1+\beta\mu_{\bf k}^2)^2b_g^2P_m(k)+\frac{1}{n_g}\right]^2\mathcal{P}_{\ell}(\mu_{\bf k})\mathcal{P}_{\ell'}(\mu_{\bf k'}).\\
\end{split}
\ee
These integrals also have analytic solutions that can be found, for example, in the Appendix of \cite{yoo/seljak:2015}. For convenience we provide them here:
\be
\begin{split}
\mathcal{C}_{0,0}(k,k)&=\frac{1}{n_g^2}+\frac{2}{n_g}\left(1+\frac{2}{3}\beta+\frac{1}{5}\beta^2\right)b_g^2P_m(k)+\left(1+\frac{4}{3}\beta+\frac{6}{5}\beta^2+\frac{4}{7}\beta^3+\frac{1}{9}\beta^4\right)b_g^4P_m^2(k)\\
\mathcal{C}_{0,2}(k,k)&=\frac{8}{n_g}\beta\left(\frac{1}{3}+\frac{1}{7}\beta\right)b_g^2P_m(k)+8\beta\left(\frac{1}{3}+\frac{3}{7}\beta+\frac{5}{21}\beta^2+\frac{5}{99}\beta^3\right)b_g^4P_m^2(k)\\
\mathcal{C}_{0,4}(k,k)&=\frac{16}{35n_g}\beta^2b_g^2P_m(k)+48\beta^2\left(\frac{1}{35}+\frac{2}{77}\beta+\frac{1}{143}\beta^2\right)b_g^4P_m^2(k)\\
\mathcal{C}_{2,2}(k,k)&=\frac{5}{n_g^2}+\frac{10}{n_g}\left(1+\frac{22}{21}\beta+\frac{3}{7}\beta^2\right)b_g^2P_m(k)+5\left(1+\frac{44}{21}\beta+\frac{18}{7}\beta^2+\frac{340}{231}\beta^3+\frac{415}{1287}\beta^4\right)b_g^4P_m^2(k)\\
\mathcal{C}_{2,4}(k,k)&=\frac{16}{7n_g}\left(3\beta+\frac{17}{11}\beta^2\right)b_g^2P_m(k)+16\beta\left(\frac{3}{7}+\frac{51}{77}\beta+\frac{435}{1001}\beta^2+\frac{15}{143}\right)b_g^4P_m^2(k)\\
\mathcal{C}_{4,4}(k,k)&=\frac{9}{n_g^2}+\frac{18}{n_g}\left(1+\frac{78}{77}\beta+\frac{1929}{5005}\beta^2\right)b_g^2P_m(k)+9\left(1+\frac{156}{77}\beta+\frac{11574}{5005}\beta^2+\frac{1308}{1001}\beta^3+\frac{711}{2431}\beta^4\right)b_g^4P_m^2(k).\\
\end{split}
\ee
For a pure galaxy sample (before interlopers), then, the multipole power spectra and covariance depend on the galaxy bias, $b_g$, galaxy number density, $n_g$, and cosmological parameters that determine $\beta$ and $P_m(k)$. Note that the sample variance and shot noise contributions to the covariance do not separate because the power spectrum covariance is a four-point function of the density field.

\section{Comparison with log-normal simulations}

Here we provide some more details of the comparison with mock galaxy catalogs from log-normal simulations using publicly-available code\footnote{\href{https://bitbucket.org/komatsu5147/lognormal\_galaxies}{https://bitbucket.org/komatsu5147/lognormal\_galaxies}} described by \cite{agrawal/etal:2017}. The code generates catalogs of three-dimensional galaxy position and velocity vectors by Poisson sampling the density field given an input matter power spectrum, galaxy bias, mean ELG density, and choice of resolution scale, using Fast Fourier Transform (FFT). We first produce two separate ELG catalogs for the target and interloper lines within cuboid volumes with side lengths specified by the sky area and redshift range for the survey under consideration. We then randomly select galaxies from the interloper line catalog to be misidentified according to the specified contamination fraction, $f_c$,  remap their three-dimensional positions as $(x_{\rm int},y_{\rm int},z_{\rm int})\to(\gamma_{\bot}x_{\rm int},\gamma_{\bot}y_{\rm int},\gamma_{\parallel}z_{\rm int})$, and add them to the catalog the target ELGs. Finally, the code computes the redshift-space multipole power spectra of the contaminated catalog.

Note that some coupling exists between multipoles of the power spectrum computed from the simulations, as described in Appendix~D.2.3 of \cite{agrawal/etal:2017}. Recovering an unbiased estimate for each multipole to compare with our analytic calculations then requires a deconvolution with a multipole mixing matrix. On large scales (small $k$) this deconvolution is not always numerically stable, which can result in fluctuations in the multipole power spectra at low $k$. An alternative option in the code is to imbed the cuboid survey volume in a larger cube, then performing FFT on the cube, with the density field set to zero outside the survey box. While this means we avoid the numerical issue associated with the multipole deconvolution, it has the downside of introducing coupling between $k$ bins. We opted to leave the $k$ bins uncorrelated, and accept some numerical instability in recovery of the large-scale power in the log-normal simulations.

The main goal of computing the simulated power spectra was to verify the analytic calculations used in our Fisher forecasts, particularly to check that each of the terms in the power spectrum covariance was being computed correctly. The green crosses in Figure~6 show the sample variance of the monopole power spectrum estimated from 500 simulations for a Euclid-like [OIII] survey with H$\alpha$ interlopers (Table~3) and $f_c=0.2$. The scatter in the points is due to the finite number of simulations, except for the second and fourth bins, where the deconvolution effect mentioned above causes larger scatter. The solid lines show the variance computed using the approximations described in Section~2.4. The contribution from the [OIII] variance is given by the first line of (A9) multiplied by $(1-f_c)^4$ and scaled by the number of modes in each $k$-bin. The contribution from the interloper H$\alpha$ ELGs is computed from equation (9) and includes the effect of coordinate remapping. Note that because the power spectrum variance is a four-point product of the density field there is also a contribution from two factors of [OIII] ELG overdensity and two H$\alpha$, labeled `[OIII]-H$\alpha$', even though the populations are independent. All of the lines shown in Figure~6 include both sample variance and shot-noise. Agreement between the simulated and calculation covariance is similar for other multipoles.

The Fisher forecasts assume that the power spectrum estimated for each bin is Gaussian distributed. We verify that this is a reasonable approximation by examining the histogram of $P_{\ell,t}(k_b)$ values from the log-normal simulations. Figure~7 shows the distribution of the first two and last two bins of the monopole power spectrum from the same 500 realizations as in Figure~6. The agreement indicates that the likelihood is sufficiently Gaussian even for the large scales where the number of modes in each bin is smallest.

\begin{figure}
\centering
\includegraphics{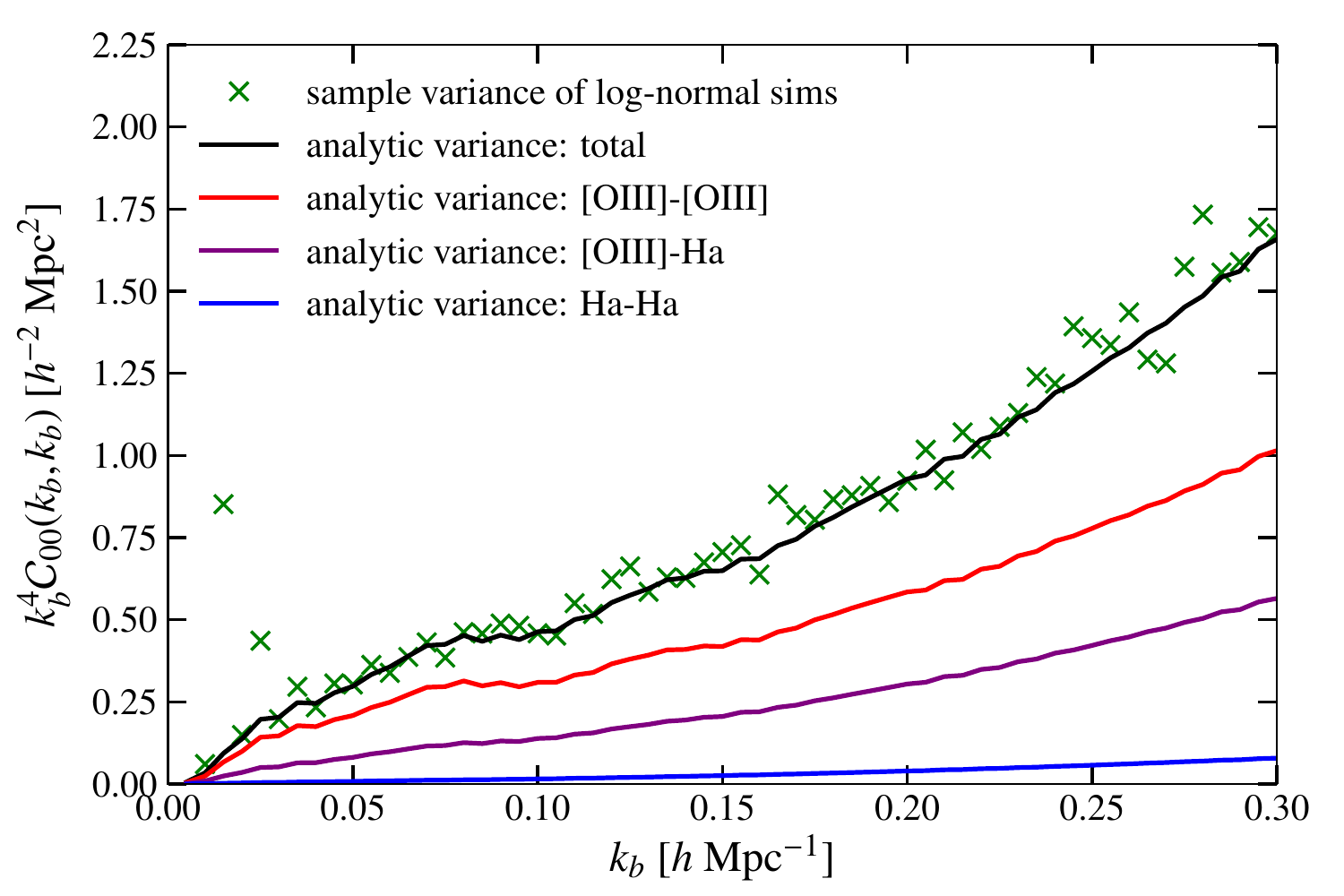}
\caption{Comparison of variance of monopole power spectrum, $\mathcal{C}_{00}$ for an [OIII] survey contaminated with H$\alpha$ with fractional contamination $f_c=0.2$. Green crosses correspond to sample variance measured in bins of width $\Delta k=0.005$~h~Mpc$^{-1}$ from 500 log-normal simulations using code described by \cite{agrawal/etal:2017} with interlopers remapped by-hand as described in the text. Solid lines show results of analytic calculations (Section~2). Consistency between the log-normal and analytic results provides a verification of the analytic calculations when interlopers are included and integrals over $\mu_{\bf k}$ need to be solved numerically. Results are similar for the variance of other multipoles and covariance between multipoles, $\mathcal{C}_{\ell\ell'}$. Large deviations in the second and fourth bins are due to numerical instability in the multipole-mixing matrix deconvolution applied to the power spectra estimated from the simulations, as explained in the text.}
\end{figure}

\begin{SCfigure}
\centering
\includegraphics{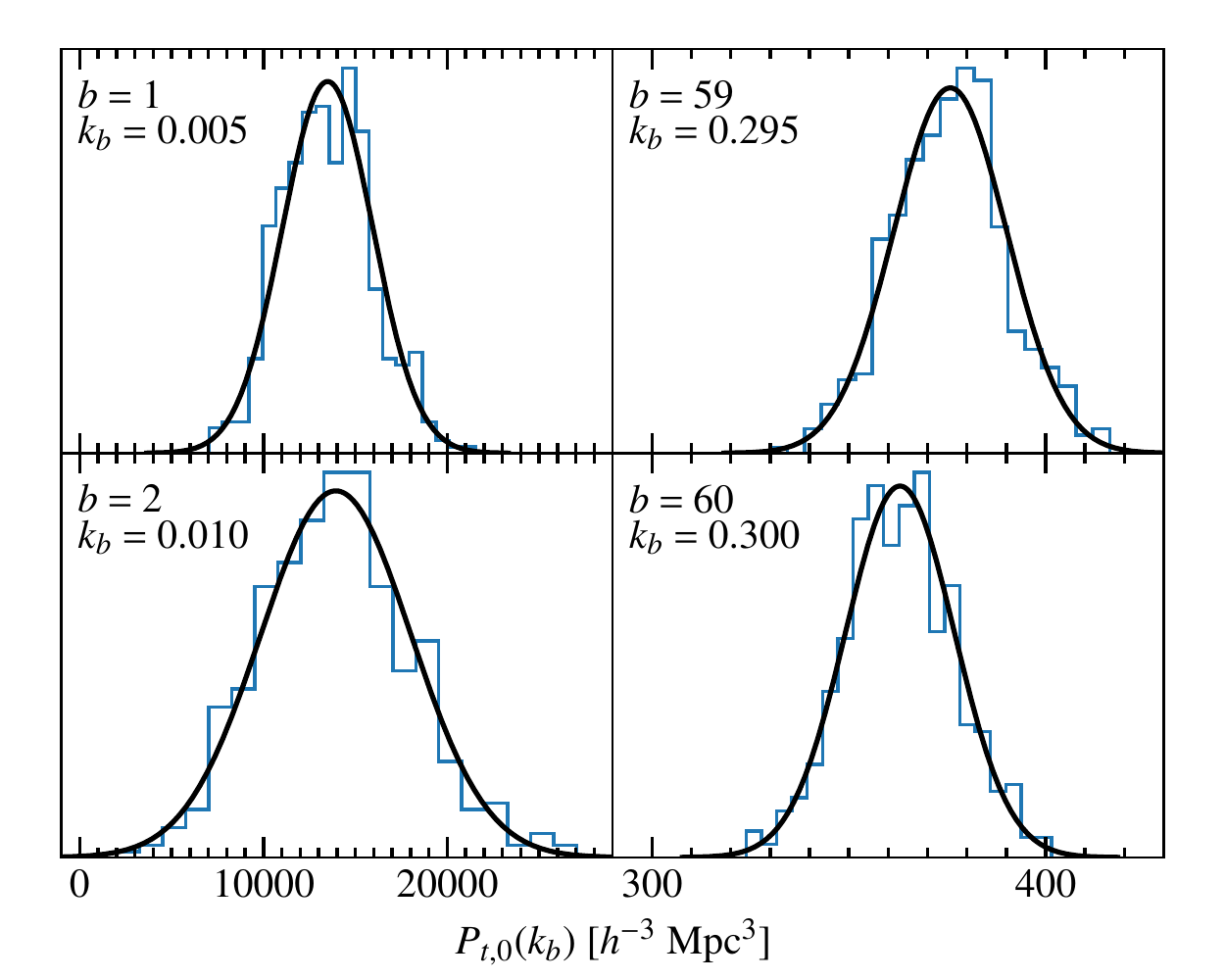}
\caption{Distribution of binned power spectra estimated from the log-normal simulations (blue histograms) is consistent with Gaussian, justifying the Gaussian likelihood approximation in Section~2.1, even though the overdensity field itself is not Gaussian. We show bins at either end of the $k$ range used in the main analysis in Section~4 ($k_b$ indicated in each panel in units of $h$~Mpc$^{-1}$). The black lines show Gaussian density function with the same mean and variance. Results are shown for simulations of the same [OIII] ELG sample contaminated by H$\alpha$ as in Figure~6 ($f_c=0.2$).}
\end{SCfigure}

\end{document}